\documentclass[12pt,preprint]{aastex}

\newcommand{\hhh}{H$_3^+$}

\slugcomment{Submitted to Astrophysical Journal}

\shorttitle{\hhh\ in the Diffuse ISM}
\shortauthors{McCall et al.}

\begin{document}


\title{Observations of \hhh\ in the Diffuse Interstellar Medium}


\author{B. J. McCall,\altaffilmark{1,2}
K. H. Hinkle,\altaffilmark{3}
T. R. Geballe,\altaffilmark{4}
G. H. Moriarty-Schieven,\altaffilmark{5}
N. J. Evans II,\altaffilmark{6}
K. Kawaguchi,\altaffilmark{7}
S. Takano,\altaffilmark{8}
V. V. Smith,\altaffilmark{9}
and T. Oka\altaffilmark{1}}


\altaffiltext{1}{Department of Chemistry, Department of Astronomy \&
Astrophysics, and the Enrico Fermi Institute, University of Chicago,
Chicago, IL 60637}
\altaffiltext{2}{Present address: Department of Astronomy, University of
California at Berkeley, 601 Campbell Hall, Berkeley, CA 94720; 
bjmccall@astro.berkeley.edu}
\altaffiltext{3}{National Optical Astronomy Observatories,
Tucson, AZ 85726}
\altaffiltext{4}{Gemini Observatory, 670 North A'ohoku Place,
Hilo, HI 96720}
\altaffiltext{5}{Joint Astronomy Centre, Hilo, HI 96720}
\altaffiltext{6}{
Department of Astronomy, University of Texas at Austin, Austin, TX 78712-1083
}
\altaffiltext{7}{
Faculty of Science, Okayama University, 3-1-1, Tsushima-naka, Okayama 700-8530, Japan
}
\altaffiltext{8}{Nobeyama Radio Observatory, 
Minamimaki, Minamisaku, Nagano, 384-1305, Japan
}
\altaffiltext{9}{
Department of Physics, University of Texas at El Paso, El Paso, TX 79968; and McDonald Observatory, University of Texas at Austin, Austin, TX 78712}


\begin{abstract}

Surprisingly large column densities of \hhh\ have been detected using infrared
absorption spectroscopy in seven diffuse cloud sightlines (Cygnus OB2 12, 
Cygnus OB2 5, HD 183143, HD 20041, WR 104, WR 118, and WR 121), demonstrating
that \hhh\ is ubiquitous in the diffuse interstellar medium.
Using the standard model of diffuse cloud chemistry, our
\hhh\ column densities imply unreasonably long path lengths ($\sim$ 1 kpc)
and low densities ($\sim$ 3 cm$^{-3}$).
Complimentary millimeter-wave, infrared, and visible observations of related 
species suggest that the chemical model is incorrect and that the number 
density of \hhh\ must be increased by one to two orders of magnitude.
Possible solutions include a reduced electron fraction,
an enhanced rate of H$_2$ ionization, and/or
a smaller value of the \hhh\ dissociative recombination rate constant than
implied by laboratory experiments.

\end{abstract}




\section{Introduction}

The \hhh\ molecular ion has long been considered to play an important
role in the chemistry of dense molecular clouds, as it initiates the network
of ion-neutral reactions \citep{herbst:klemp,wdwatson} that is responsible for 
the wealth of molecules observed by infrared and radio astronomers.  However,
\hhh\ has not been considered to be an important species in diffuse clouds,
because it is thought to be destroyed rapidly by dissociative recombination
with electrons, which are abundant in diffuse clouds.
For the purposes of the present study, we do not make a distinction between
diffuse and translucent clouds, but simply use ``diffuse'' to indicate
sightlines where CO is not the dominant reservoir of carbon (in contrast to
dense clouds).

The detection of \hhh\ in the diffuse interstellar medium toward the
Galactic Center \citep{geballediffuse} and toward the visible star
Cygnus OB2 12 \citep{mccalldiffuse} with similar column densities to those of
dense clouds \citep{densecloud} was therefore quite surprising.  These
observational results imply either a very long pathlength (hundreds of
parsecs) of absorbing material, or a serious problem with the standard
model of diffuse cloud chemistry.
Various attempts have been made to explain the abundance of \hhh\
toward Cygnus OB2 12 while preserving the standard model of the chemistry,
either by adjusting all parameters to optimize \hhh\ \citep{cp:dalgarno}
or by postulating an additional ionization source for H$_2$ molecules
peculiar to the
neighborhood of the Cygnus OB2 association \citep{black:rs}.

We have conducted a small survey of twelve diffuse cloud sources with the
aim of determining whether the sightline toward Cygnus OB2 12 is unique,
or whether high column densities of \hhh\ are common in diffuse clouds.
As a result of this survey, we have detected \hhh\ in seven diffuse
cloud sources (Cygnus OB2 12, Cygnus OB2 5, HD 183143, HD 20041,
WR 104, WR 118, and WR 121), and obtained upper limits toward five other 
sources (HD 194279, HD 168607, P Cygni, $\chi^2$ Ori,
and $\zeta$ Oph).  For Cygnus OB2 12 and 5 and HD 181343, we have also
obtained infrared and radio spectra of CO, as well as high-resolution
visible spectra of relevant atoms and molecules.  We have also obtained
infrared CO spectra towards WR 104 and WR 108.

These results demonstrate the ubiquity of \hhh\ in diffuse clouds, and 
suggest that there is a ``global'' problem with the current models.
In this respect, \hhh\ represents a major problem in diffuse
cloud chemistry, reminiscent of the Diffuse Interstellar Bands and CH$^+$.

\section{Observations and Data Reduction}

The stellar parameters of our targets are listed in Table \ref{targets}, and
a summary of the observations is provided in Table \ref{obslog}.
The $R$(1,1)$^l$ line of \hhh\ 
at 3.715 $\mu$m was observed using
the Phoenix spectrometer on the Mayall 4-meter telescope at Kitt Peak
National Observatory (KPNO).  The \hhh\ observations were performed in three
runs (July 1998, June 2000, and March 2001).  
The CGS4 facility spectrometer on the United
Kingdom Infrared Telescope (UKIRT) was used in July 2000 and May 2001 
to study the $R$(1,0)--$R$(1,1)$^u$ doublet of \hhh\ near 3.668 $\mu$m.  
The three observed \hhh\ rotation-vibration transitions belong to the 
fundamental band of the $\nu_2$ degenerate bending mode, and arise from 
the lowest allowed rotational states ($J$,$K$) = (1,1) and (1,0).  
The old notation [e.g., $R$(1,1)$^+$] has been superseded by a more
flexible notation [e.g., $R$(1,1)$^u$] which uses $u$ and $l$ in place
of $+$ and $-$ [for a complete description of the new standard
notation for \hhh, see \citet{myrs} or \citet{myjcp}].

Infrared spectra of portions of 
the fundamental ($v$=1-0) and overtone ($v$=2-0) 
infrared bands of CO were also obtained.  The fundamental band was
studied toward Cygnus OB2 12 and Cygnus OB2 5 with Phoenix in July 1999
and June 2000, and toward HD 183143, WR 104, and WR 118 with CGS4 in
May 2001.  The overtone band of CO was searched for with Phoenix
toward HD 183143 in June 1997 and toward Cygnus OB2 12 in July 1999.

All the infrared data were reduced using
the procedure outlined in \citet{densecloud}, which is described in
more detail in \citet{mythesis}.  Object spectra were ratioed by
(scaled) standard spectra in order to remove the many strong atmospheric
absorption lines in these regions.  The wavelength calibration
was achieved using telluric absorption lines, and is estimated to be accurate
to roughly 2 km/s.  The resolving power of the Phoenix spectrometer was 
roughly 40,000 until 1998, and approximately 60,000 during 1999--2001.  The
CGS4 spectrometer was used with the long camera, yielding a resolving power
of 40,000.

High resolution visible spectra of atoms and molecules of interest
were obtained using the Coud\'{e} spectrometer on the Smith 2.7-m telescope
at McDonald Observatory.  A total of three echelle settings were used.  The
first covered the K \textsc{i} line at 7699 \AA\ and the CN A-X $v$=2-0 $R_2(0)$ line
at 7875 \AA.  The second covered the C$_2$ A-X $v$=2-0 band near 8760 \AA.
These two echelle settings were used with the F1E2 configuration and the
0.2'' slit, which yielded a resolving power of about 200,000.  The third
echelle setting covered Ca \textsc{i} at 4227 \AA, Ca \textsc{ii} 
at 3934 \AA, CH at 4300 \AA,
CH$^+$ at 4232 \AA, and CN B-X $v$=0-0 near 3874 \AA.  This setting was used
with the F1E1 configuration and the 0.6'' slit, to yield a resolving power
of approximately 120,000.  The McDonald data were reduced using standard
IRAF routines (\textsc{ccdproc} and \textsc{doecslit}).  It was not
generally necessary to ratio the visible spectra by standard stars, due to
the weakness and paucity of atmospheric absorption lines in the regions of
interest.  Wavelength calibration was achieved using comparison
spectra of a Th-Ar lamp, and is estimated to be good to about 0.01 \AA\ in
all cases.

Rotational spectra of CO in the Cygnus OB2 association 
were obtained at the Nobeyama Radio Observatory (NRO) for $J$=1-0,
and the Caltech Submillimeter Observatory (CSO) for $J$=2-1 and 3-2.  
The CSO spectra were obtained using position switching with an ``off''
position of 30' south.  This position has significant $^{12}$CO emission,
but no detectable $^{13}$CO emission compared with a more distant off position,
which is important as the $^{13}$CO data is used in the interpretation of the
infrared CO lines.
The NRO spectra were obtained using position switching with an off position
of $(l=81,b=3)$, which was found to be more free of CO emission.
CO spectra toward HD 183143 were also obtained at
the James Clerk Maxwell Telescope (JCMT) for $J$=2-1, and at NRO for
$J$=1-0, using frequency switching.  In all cases, data were reduced
using the standard procedures.

\section{Results}
\subsection{\hhh\ Spectra}

\begin{figure}
\begin{center}
\epsscale{0.54}
\plotone{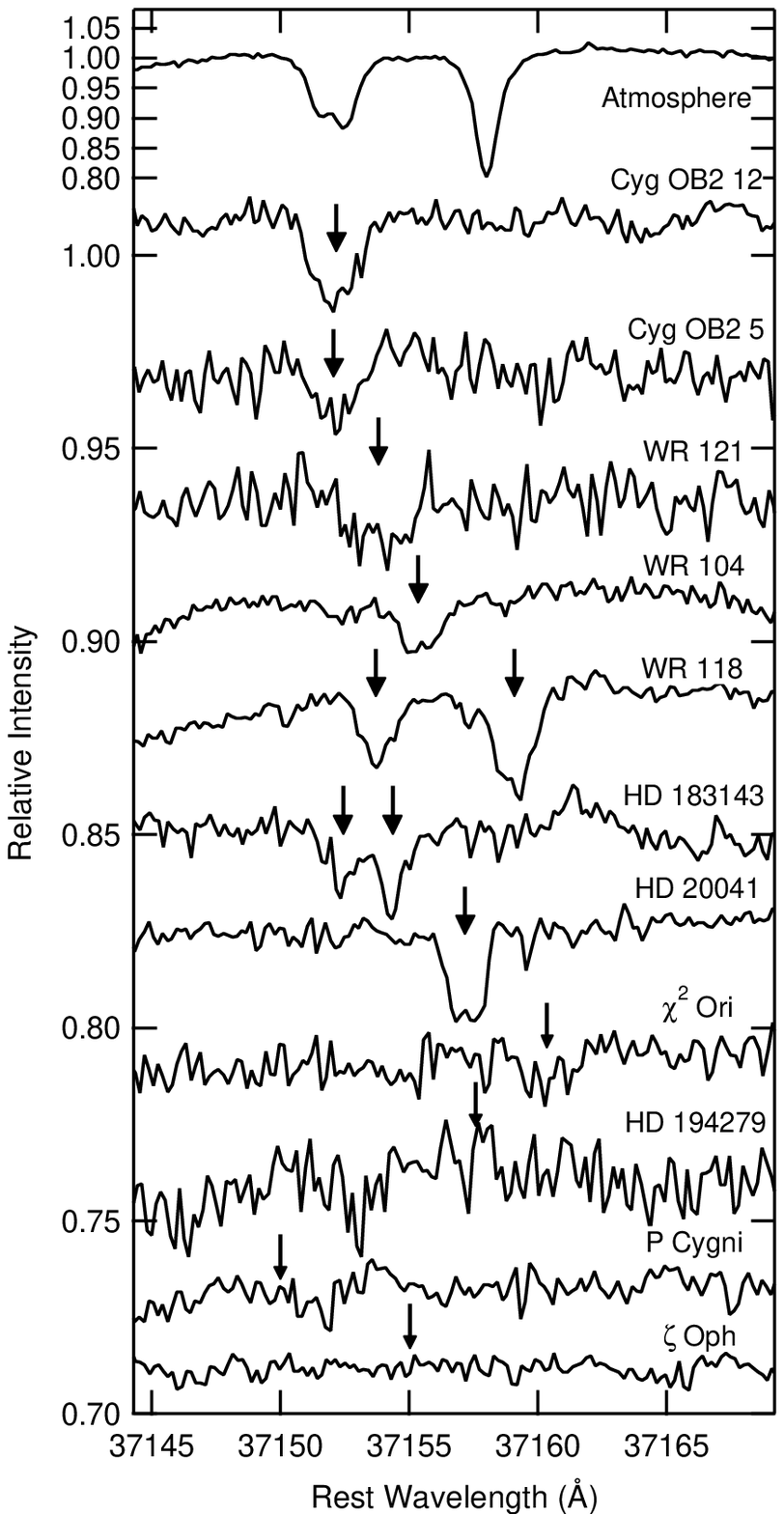}
\caption{Spectra of the \hhh\ $R$(1,1)$^l$ line observed with Phoenix at
Kitt Peak.  
Each object spectrum has been ratioed by that of a standard star
to remove atmospheric absorption lines.
The vertical arrows denote the observed (thick arrows) or 
expected (thin arrows) positions of \hhh\ depending on each source's
Doppler shift at the time of observation.\label{KPNO_H3plus}}
\end{center}
\end{figure}

In our 1998 observing run at Kitt Peak, we detected \hhh\ for the first
time toward Cygnus OB2 5, which lies approximately 6' (3 pc at 1.7 kpc)
away from Cygnus OB2 12.  Although the \hhh\ $R$(1,1)$^l$ line was 
unfortunately
Doppler shifted under the telluric CH$_4$ feature near 37152 \AA, we are
confident that the detection is secure, because the other telluric CH$_4$ line
(near 37158 \AA) is removed very effectively by ratioing with a standard
star, and because we observed a similar
\hhh\ feature in Cygnus OB2 12 at the same velocity with an equivalent
width consistent with our earlier observations \citep{mccalldiffuse}.
Our Kitt Peak \hhh\ observations are summarized in Figure \ref{KPNO_H3plus}.

Our 1998 run also yielded detections of \hhh\ toward the Wolf-Rayet
stars WR 121, WR 104, and WR 118.  The source WR 118 is interesting in that
it shows two \hhh\ components, which have similar (though not the
same) spacing as the two telluric CH$_4$ lines.  Consequently, the scaling and
ratioing by the standard star was performed by choosing parameters which 
minimized the residuals of CH$_4$ lines elsewhere in the spectrum (not
shown in the figure).  It is worth noting that the inferred velocities of
the peaks ($\sim$+5 and +48 km/s) would occur at heliocentric distances
of $\sim$0.5 and 3 kpc, respectively, assuming a flat Galactic rotation 
curve with $v_c$ = 220 km/s, $R_0$ = 8 kpc, and $l$ = 21$^{\circ}$.80.
These distances do not exceed the estimated distance to the source
($m-M$ = 14, or
$d \sim$ 6.3 kpc) obtained from the observed V magnitude of 22, the visual
extinction $A_V$ = 12.8 from \citet{vanderHucht}, and the absolute magnitude
$M_V$ = -4.8 for WC9 stars \citep{wolfrayet}.

Finally, our 1998 Kitt Peak run yielded a non-detection toward HD 194279,
toward which CH is observed with a heliocentric velocity of -11.8 km/s
(G.\ Galazutdinov, private communication).  This source, along with the
Wolf-Rayet stars, was chosen for study because the 3.4 $\mu$m aliphatic
carbon feature (a signature of diffuse clouds)
had been detected by \citet{pendleton}.

Our June 2000 run at Kitt Peak yielded the first detection of \hhh\ toward
a more traditional diffuse cloud source, HD 183143.  Two components of \hhh\
were seen, in agreement with the visible spectra of other molecules (see
below), although the blue component was somewhat affected by the telluric
CH$_4$ line.  During this run, we also obtained non-detections toward
P Cygni [which has an LSR velocity of CH$^+$ of -9 km/s \citep{hobbs73}]
and toward $\zeta$ Oph [which has an LSR velocity of CO of -0.79 km/s
\citep{lambert90}].
Our March 2001 run at Kitt Peak yielded a detection of \hhh\ toward the least
reddened star to date, HD 20041.  We also obtained an upper limit of \hhh\
absorption towards $\chi^2$ Ori.

\begin{figure}
\begin{center}
\epsscale{0.54}
\plotone{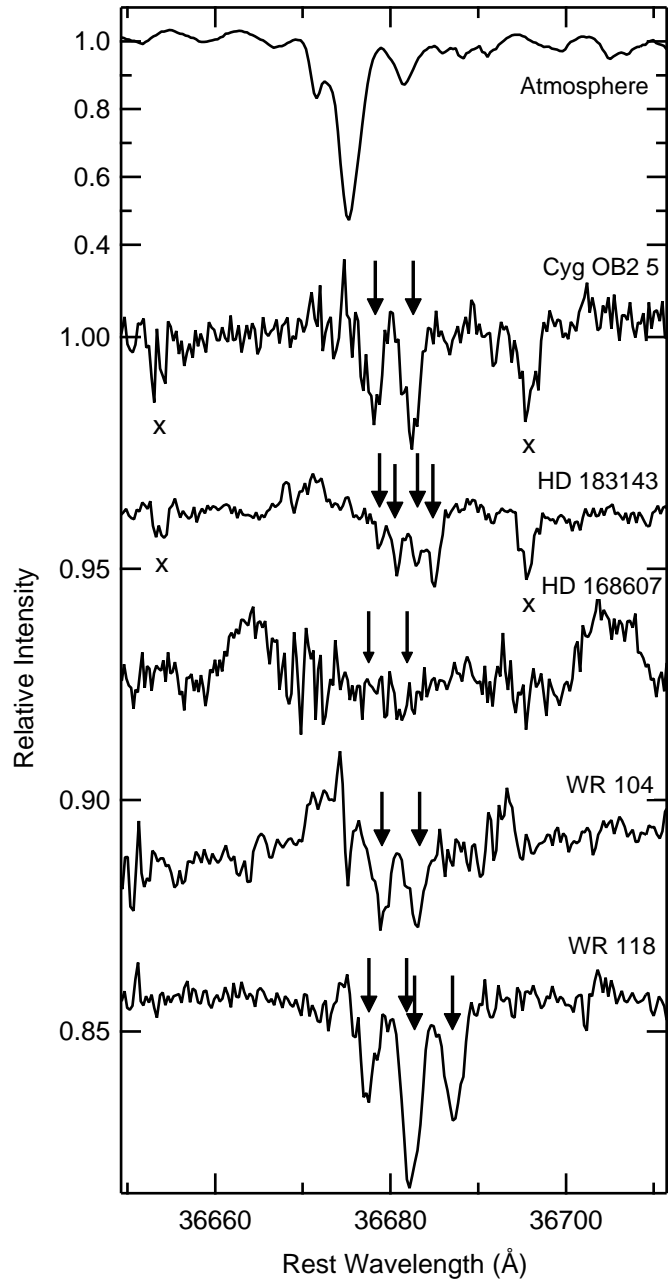}
\caption{Spectra of the \hhh\ $R$(1,0)--$R$(1,1)$^u$ doublet of \hhh\
observed with CGS4 at UKIRT.  
Each object spectrum has been ratioed by that of a standard star
to remove atmospheric absorption lines.
The vertical arrows denote the observed or
expected (for HD 168607) position of the \hhh\ doublet.  The crosses label 
instrumental artifacts.
\label{UKIRT_H3plus}}
\end{center}
\end{figure}

Our run at UKIRT in July 2000 (see Figure \ref{UKIRT_H3plus}) confirmed
the detections of \hhh\ toward Cygnus OB2 5 and HD 183143.  In the case
of HD 183143, the two-component velocity structure is very clear.  For
both the Cygnus OB2 5 and HD 183143 observations, instrumental artifacts
were present in the ``negative'' spectrum --- these features changed in
intensity (and sometimes disappeared) depending on exactly which rows of
the array were extracted, while the \hhh\ features were consistently
present.  We
also obtained a non-detection toward HD 168607 [which has a heliocentric
CH velocity of 22 km/s \citep{gredel99}].  

In May 2001 at UKIRT, we confirmed the detections of \hhh\ toward the
Wolf-Rayet stars WR 104 and WR 108 by using the $R$(1,1)$^u$--R(1,0)
doublet.  The separation of the two velocity components toward WR 118 is
comparable to the spacing between the two \hhh\ transitions, so that
the $R$(1,0) of the blue component and the $R$(1,1)$^u$ of the red
component are overlapped.

A table of the observed \hhh\ line parameters is given in 
Table \ref{h3pluslines}.  The observed equivalent widths have been
converted to column densities under the assumption that the lines
are optically thin, using the transition dipole moments calculated by
J.~K.~G.~Watson, as listed in \cite{mccalldiffuse}.

\subsection{CO towards the Cygnus OB2 Association}
\label{cocyg12}

\begin{figure}
\begin{center}
\epsscale{1}
\plotone{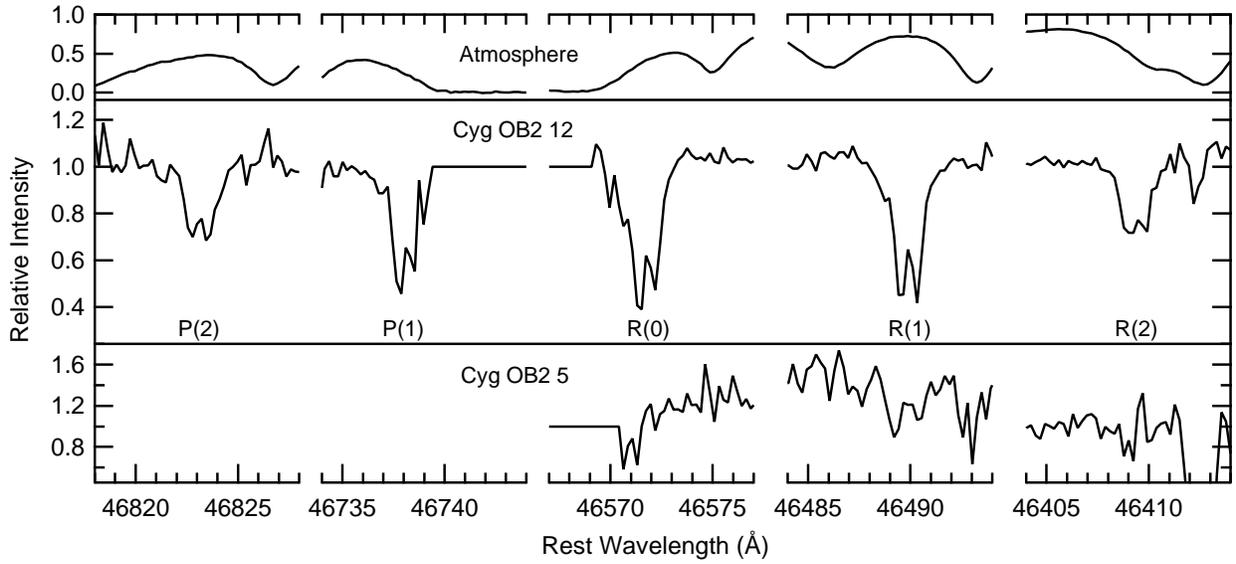}
\caption{Infrared $v$=1-0 CO spectra obtained with Phoenix at Kitt Peak.
Object spectra have been ratioed by standard star spectra to remove the
atmospheric CO lines.  The horizontal segments of the spectra indicate
regions where the strong telluric absorptions could not be ratioed out.
\label{COIR}}
\end{center}
\end{figure}

A brief look at the high resolution infrared spectrum of CO  toward 
Cygnus OB2 12 and Cygnus OB2 5 was obtained during test and engineering time 
with Phoenix at Kitt Peak in July 1999.  Followup observations with longer
integration time in June 2000 yielded higher signal-to-noise spectra,
displayed in Figure \ref{COIR}.  The observations of Cygnus OB2 12 clearly 
show the two
closely-spaced ($\sim$ 5 km/s) velocity components observed in the mm-wave
spectrum from JCMT \citep{geballediffuse}.  
The infrared lines suggest a CO excitation temperature of $\sim$10 K
toward Cygnus OB2 12, but this should not be taken as an indication of the 
kinetic temperature of the gas due to the effect of spontaneous
emission \citep{geballediffuse}.  The infrared CO line parameters are listed
in Table \ref{ircolines}.  We have adopted the transition dipole moments of
\citet{hure}.

\begin{figure}
\begin{center}
\epsscale{0.8}
\plotone{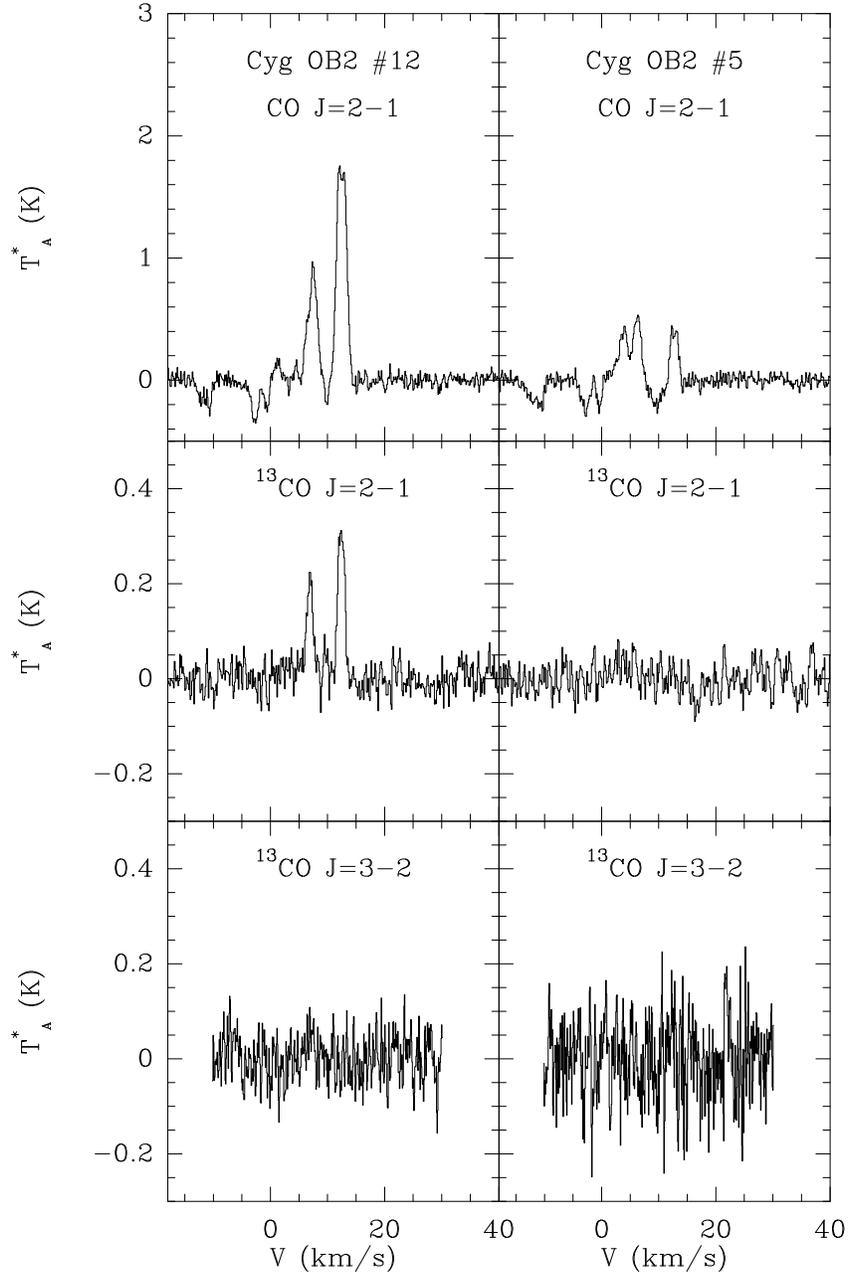}
\caption{CO spectra obtained at CSO.
\label{neal}}
\end{center}
\end{figure}

The signal-to-noise ratio of the Cygnus OB2 12 spectrum is clearly
not high enough to rely on the measured equivalent widths of the individual
components, particularly because of the contamination of the telluric lines.
However, we can estimate the column density of CO by assuming that the true
equivalent widths of the two components have the same ratio as the 
integrated areas of the components in the $^{13}$CO emission spectrum, and 
by adopting the $b$-values of the $^{13}$CO in estimating the saturation
corrections.  This analysis was performed with the high-quality $^{13}$CO
spectrum obtained at CSO (shown in Figure \ref{neal}; the mm-wave line
parameters are listed in Table \ref{mmcolines}).  We assume that the 
7 km/s peak carries 0.39 of the equivalent width (and has $b$=0.75 km/s), 
and that the 12 km/s peak carries 0.61 of the equivalent width (and has
$b$=0.88 km/s).  With these assumptions we estimate a total CO column
density of $\sim 1.4 \times 10^{17}$ cm$^{-2}$ in front of Cygnus OB2 12.
This estimate is considerably
higher than the estimate of $3 \times 10^{16}$ cm$^{-2}$ given in
\citet{geballediffuse}, but still much lower than the total column
density of carbon ($\sim 2.5 \times 10^{18}$ cm$^{-2}$) inferred from the
color excess.  

We have also obtained an upper limit on the CO column density toward 
Cygnus OB2 12, based on our failure to detect the
$v$=2-0 overtone with Phoenix.  The upper limit on the $v$=2-0 features yield 
an upper limit (3$\sigma$) of 
N(CO) $\stackrel{<}{_{\sim}} 3 \times 10^{17}$ cm$^{-2}$.  This suggests
that at most about 10\% of the carbon along the line of sight to Cygnus OB2
12 can be in the form of CO, and that this sightline does not consist of
dense clouds.

\begin{figure}
\begin{center}
\epsscale{0.54}
\plotone{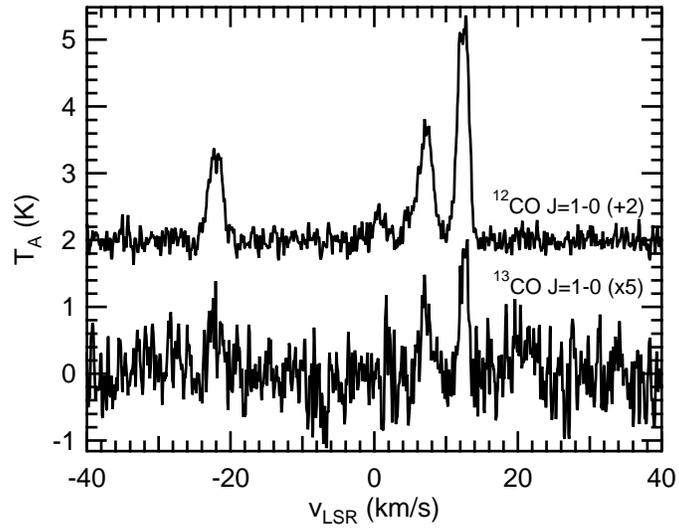}
\caption{CO $J$=1-0 spectra of Cygnus OB2 12 obtained at NRO.
\label{Nobeyama12}}
\end{center}
\end{figure}

Using the $^{13}$CO data from CSO for $J$=2-1 and 
from NRO for $J$=1-0 (see Figure \ref{Nobeyama12}), we performed an
analysis using an LVG code to calculate the statistical equilibrium,
assuming a temperature of about 30 K (consistent with the C$_2$ analysis
of \S \ref{C2} and the \hhh\ analysis of \S \ref{inferred}).
If the density ($n = 200$ cm$^{-3}$) derived from the analysis of C$_2$ 
toward Cygnus OB2 12 is assumed,
the $^{13}$CO column density, $N(13)$ is determined uniquely by matching
the observed line emission. The results are 
$N(13) = 1.8 \times 10^{15}$ cm$^{-2}$ for the 7 km/s component and
$N(13) = 2.7 \times 10^{15}$ cm$^{-2}$ for the 12 km/s component.
The sum of these, $N(13) = 4.5 \times 10^{15}$ cm$^{-2}$, would imply
a total column density of CO of 2.3--$4.5 \times 10^{17}$ cm$^{-2}$
for isotope ratios of 50--100; this column density is about twice that
inferred from the infrared CO absorption lines. 
If, on the other hand, we enforce agreement with the column density from
the infrared absorption, and assume an isotope ratio of 100, the 
$^{13}$CO data indicate somewhat higher densities: 
$n = 500$ cm$^{-3}$ for the 7 km/s component, and 
$n = 900$ cm$^{-3}$ for the 12 km/s component. 
While these densities are somewhat higher than inferred from C$_2$,
it would not be surprising to have density variations, with the 
$^{13}$CO data probing the denser gas, and the C$_2$ probing somewhat
less dense gas.

The line of sight toward Cygnus OB2 5 appears to have considerably less
CO than that toward Cygnus OB2 12, based both on the marginal detections
in the infrared spectrum (Figure \ref{COIR}) as well as the CSO $^{13}$CO
spectra (Figure \ref{neal}).  Therefore we expect that this sightline is
also dominated by diffuse cloud material, although because of the low
signal-to-noise of the infrared spectrum it is difficult to estimate the
CO column density.

\begin{figure}
\begin{center}
\epsscale{0.54}
\plotone{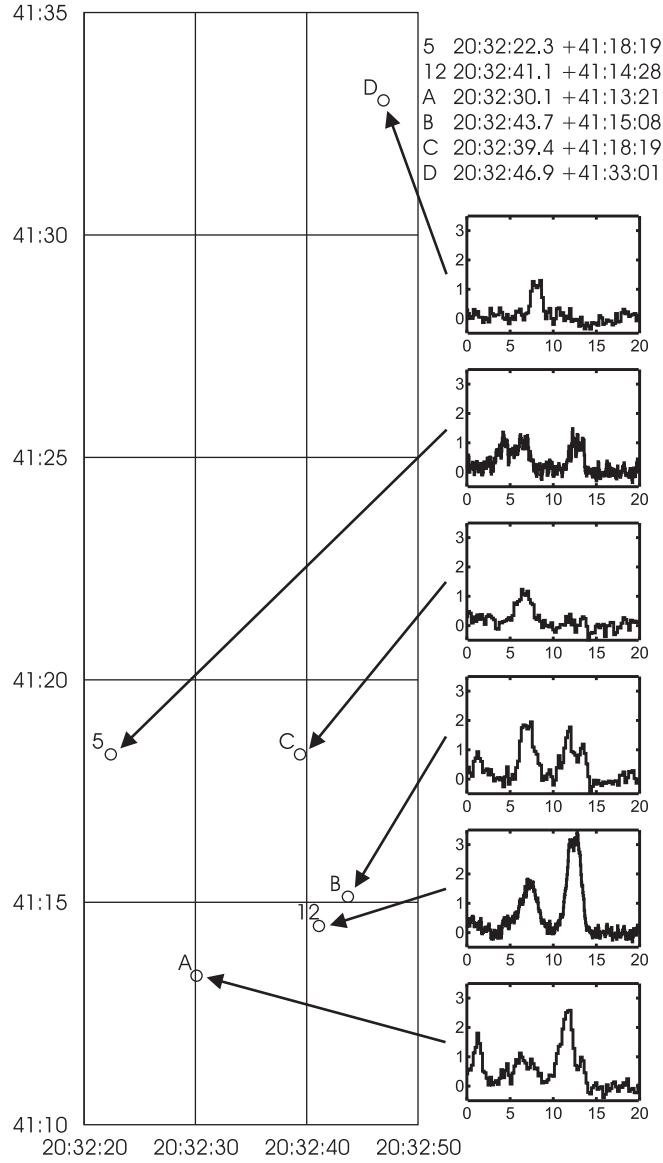}
\caption{CO $J$=1-0 ``map'' of the Cygnus OB2 association obtained at NRO.
Spectra have T$_A^*$ (K) as vertical axis, v$_{\rm LSR}$ (km/s) as horizontal
axis.  All coordinates are J2000.
\label{cygmap}}
\end{center}
\end{figure}

In order to probe the spatial extent of the CO gas in the Cygnus OB2
association, we have obtained (position-switched) 
$^{12}$CO $J$=1-0 spectra at four other locations
in the association, labelled A through D in Table \ref{obslog} and
Figure \ref{cygmap}, which also shows the spectra toward Cygnus OB2 12 and 5.

\subsection{CO towards HD 183143}

\begin{figure}
\begin{center}
\epsscale{0.54}
\plotone{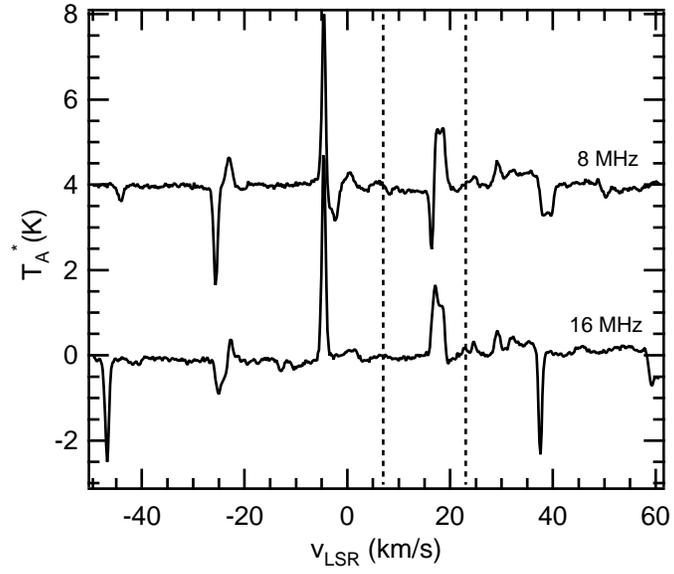}
\caption{$^{12}$CO $J$=2-1 frequency switched spectra toward HD 183143, 
obtained at JCMT.  The velocities of the \hhh\ lines are marked with 
vertical dashed lines.
\label{JCMT}}
\end{center}
\end{figure}

\begin{figure}
\begin{center}
\epsscale{0.54}
\plotone{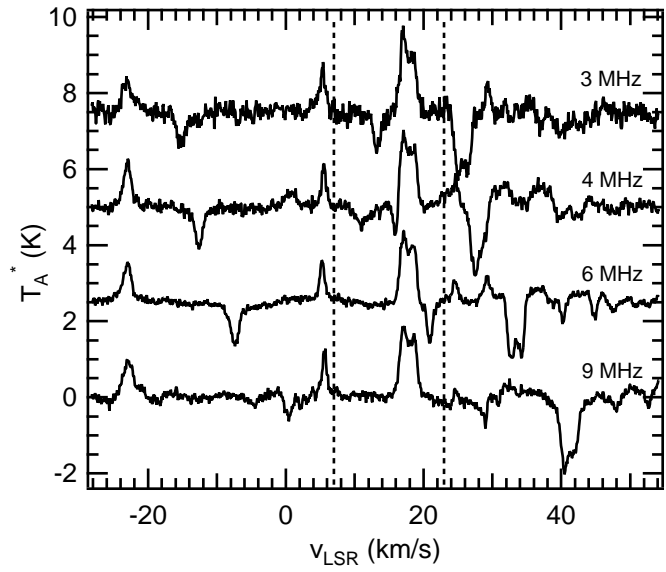}
\caption{$^{12}$CO $J$=1-0 frequency switched spectra toward HD 183143, 
obtained at Nobeyama (folding has not been performed).  The velocities 
of the \hhh\ lines are marked with vertical dashed lines.
\label{NRO183143}}
\end{center}
\end{figure}

To check for the presence of CO in this sightline, we 
obtained rotational spectra of the $J$=2-1 line (at JCMT) and the $J$=1-0
line (at NRO).  The JCMT results (frequency-switched at 8 MHz and 16 MHz)
are shown in Figure \ref{JCMT}.  While the $J$=2-1 spectrum of 
Cygnus OB2 12 showed nearly 2 K of emission at the \hhh\ velocity, the 
spectrum of HD 183143 shows no emission at the correct velocities in excess of
$\sim$ 0.1 K.  The NRO results are shown in Figure \ref{NRO183143},
and provide an upper limit of $\sim$ 0.2 K for the $J$=1-0 emission
at the observed \hhh\ velocities.  The closest reasonable velocity 
component is at +25 km/s, and is only about
0.5 K.  In contrast, Cygnus OB2 12 shows 3 K of emission.

\begin{figure}
\begin{center}
\epsscale{1}
\plotone{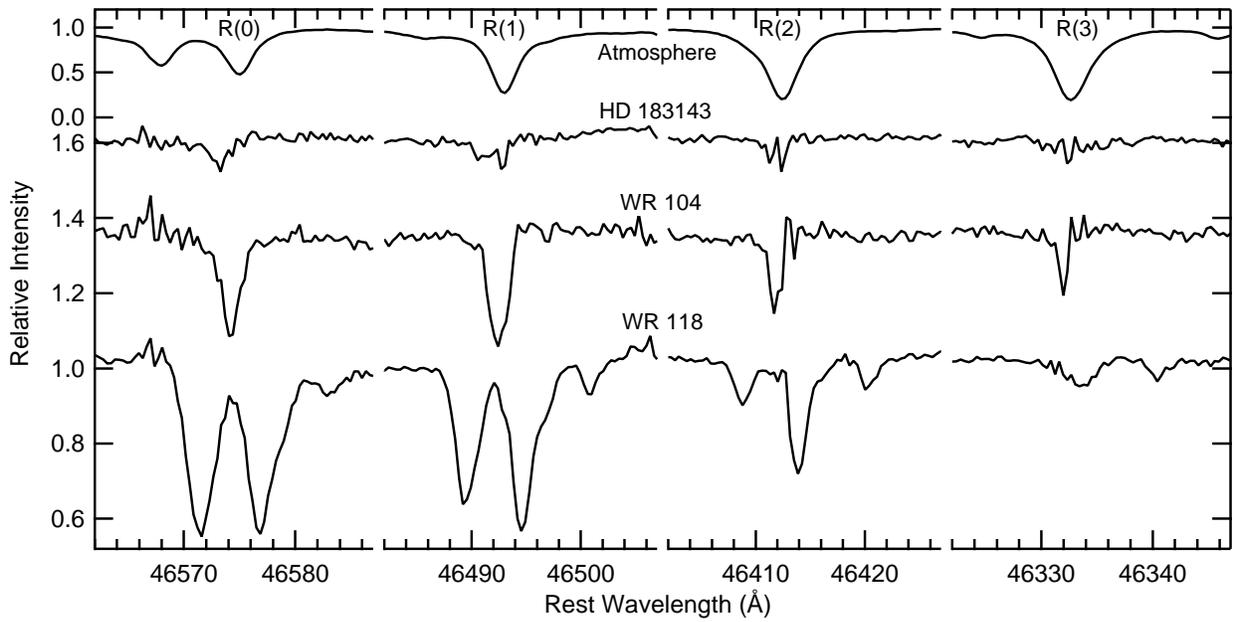}
\caption{$^{12}$CO $v$=1-0 fundamental band observations of HD 183143,
WR 104, and WR 118, obtained at UKIRT.
Object spectra have been ratioed by standard star spectra to remove the
atmospheric CO lines.
\label{newCO}}
\end{center}
\end{figure}

During our May 2001 UKIRT run, we obtained an absorption spectrum of the
$R$-branch of the CO $v$=1-0 fundamental band.  This spectrum (see Figure
\ref{newCO} and Table \ref{ircolines}) 
shows weak absorptions near $v_{\rm{LSR}} \sim$ 25 km/s,
in agreement with the radio spectra.  Because of the fairly small 
(geocentric) Doppler shift ($\sim$ 11 km/s) at the time of the observation,
it is difficult to accurately measure the column density, but it is clear
that the total CO column density is less than $10^{16}$ cm$^{-2}$.
A previously obtained Phoenix spectrum of the $v$=2-0 overtone band 
(from June 1997) provides a less stringent upper limit on the total CO 
column density of $\sim 10^{17}$ cm$^{-2}$.

Both the radio and infrared results demonstrate that there is very little 
(if any) CO associated with the \hhh\ toward HD 183143.  

\subsection{CO towards WR 104 and WR 118}

At UKIRT in May 2001, we obtained spectra of the CO fundamental band toward
the Wolf-Rayet stars WR 104 and WR 118.  The spectra are shown in Figure
\ref{newCO} and the line parameters are listed in Table \ref{ircolines}.
WR 104 shows a single velocity component centered at $v_{\rm{LSR}} \sim$
21 km/s --- surprisingly, this is different from the velocity of the \hhh\
($\sim$ 10 km/s)!  WR 118 has a very complicated velocity structure,
which we have fit with four Gaussian components.  The lowest velocity
component at $v_{\rm{LSR}} \sim$ 10 km/s may be associated with the \hhh\
component at $v_{\rm{LSR}} \sim$ 5 km/s, and the blended components near
$v_{\rm{LSR}} \sim$ 40--55 km/s are consistent with the second \hhh\
component at $v_{\rm{LSR}} \sim$ 48 km/s.

Without any radio spectra of these sources, it is difficult to attempt
saturation corrections, so we have simply listed in Table \ref{ircolines}
the column densities in the limit that the lines are optically thin.  In
this limit, WR 104 has N(CO) $\sim 9 \times 10^{15}$ cm$^{-2}$ and
WR 118 has N(CO) $\sim 4 \times 10^{16}$ cm$^{-2}$ (in the $J$=0--3 levels).

\subsection{C$_2$ Spectra}
\label{C2}

\begin{figure}
\begin{center}
\epsscale{1}
\plotone{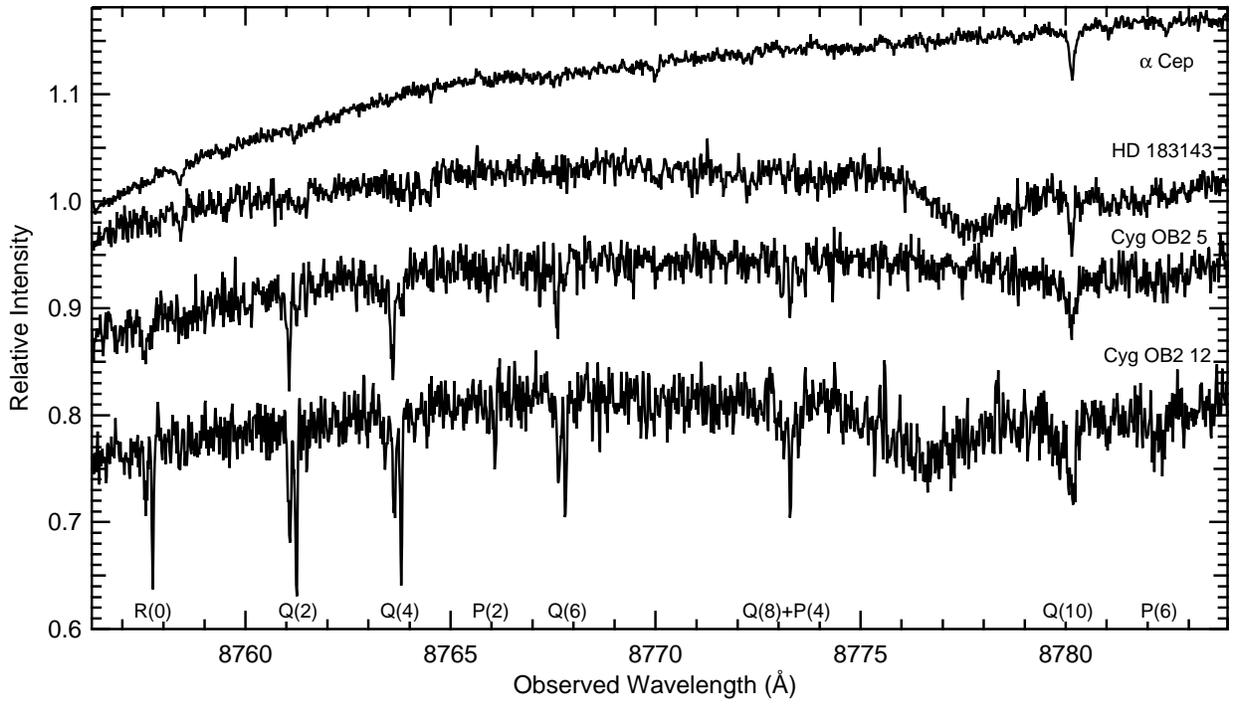}
\caption{Spectra of the A-X $v$=2-0 band of C$_2$ obtained at McDonald.
\label{C2spectra}}
\end{center}
\end{figure}

The C$_2$ A-X $v$=2-0 band was clearly detected toward Cygnus OB2 12 at
McDonald, as shown in Figure \ref{C2spectra}.  Each line appears as a
clear doublet with a separation comparable to that of the CO, except for
the $Q$(8) line which is blended with $P$(4).  The equivalent widths were
measured separately for the two velocity components, except for the $Q$(8)
+ $P$(4) blend, for which only the total equivalent width could be measured.
The contribution of $P$(4) was estimated from the strength of $Q$(4), which
permitted an estimate of the $Q$(8) equivalent width for the sum of the
two velocity components.  
The $Q$(10) line was contaminated by an atmospheric H$_2$O absorption
line, and the spectra were ratioed by that of $\alpha$ Cep in order to remove
the contribution of the telluric line. 
The derived C$_2$ line parameters (for each component separately, as well
as for the total profile) for Cygnus OB2 12 are listed in 
Table \ref{C2table}.  Column densities have been determined assuming the
lines are optically thin, using the oscillator strength $f_{20} = 1.67 \times
10^{-3}$ of \citet{EvD:f}.

Using the method of \citet{EvD:C2}, and assuming
the scaling factor for the radiation field $I=1$ and the C$_2$--H$_2$
collisional cross-section $\sigma_0 = 2 \times 10^{-16}$ cm$^2$, we can
estimate the number density $n$ of collision partners, as well as the
kinetic temperature $T$.  This analysis is performed by calculating the
rotational distribution for a grid of points on the ($n$,$T$) plane, then
finding the best fit to the column densities $N$($J$) derived from the
observed spectrum.  The advantage of this method is that it yields the
range of values of ($n$,$T$) that are consistent with the observations,
given the uncertainties in the measurements.  A web-based 
``C$_2$ calculator'' is available at http://dib.uchicago.edu/c2

For the 7 km/s component, we obtain a best
fit value of ($n$,$T$)=(220 cm$^{-3}$, 40 K), and for the 12 km/s component
we obtain (210 cm$^{-3}$, 30 K).  Using the total column densities rather
than those of the individual components yields (200 cm$^{-3}$, 30 K).
These values are fairly well constrained by the observational data ---
it is exceedingly unlikely (i.e.\ it requires more than a 3$\sigma$ error
in at least one of the measurements) that the density is outside the
range 150--600 cm$^{-3}$ or that the temperature is outside the range 25--55 K.

The spectrum of Cygnus OB2 5 also shows a hint of the doublet structure,
but the signal-to-noise of the spectrum is not sufficient to reliably
measure the equivalent widths of the individual components.  Therefore
we have measured only the total equivalent widths, which are reported
in Table \ref{C25table}.  An analysis of the rotational excitation
(while very uncertain) suggests that the density exceeds 800 cm$^{-3}$
and that the temperature lies in the range 60--90 K.  According to
\citet{gredel:munch}, chemical models predict a sudden
decrease in C$_2$ abundance at densities above 10$^{3.5}$ cm$^{-3}$.  Taken
with our (uncertain) results, this suggests that the density in the region
where C$_2$ is found is in the range 800--3000 cm$^{-3}$.

In agreement with the results of \citet{gredel99}, we saw absolutely
no trace of C$_2$ towards HD 183143.  Assuming the $Q$(2) line would have
a width of $\sim$ 5 km/s, the 3$\sigma$ limit on the equivalent width is
about 1.7 m\AA, which corresponds to a column density of N($J$=2)
$\stackrel{<}{_{\sim}} 3 \times 10^{12}$ cm$^{-2}$ (comparable to Gredel's
upper limit), which is more than ten times less than that of Cygnus OB2 12!

\subsection{Other Visible Spectra}

\begin{figure}
\begin{center}
\epsscale{0.54}
\plotone{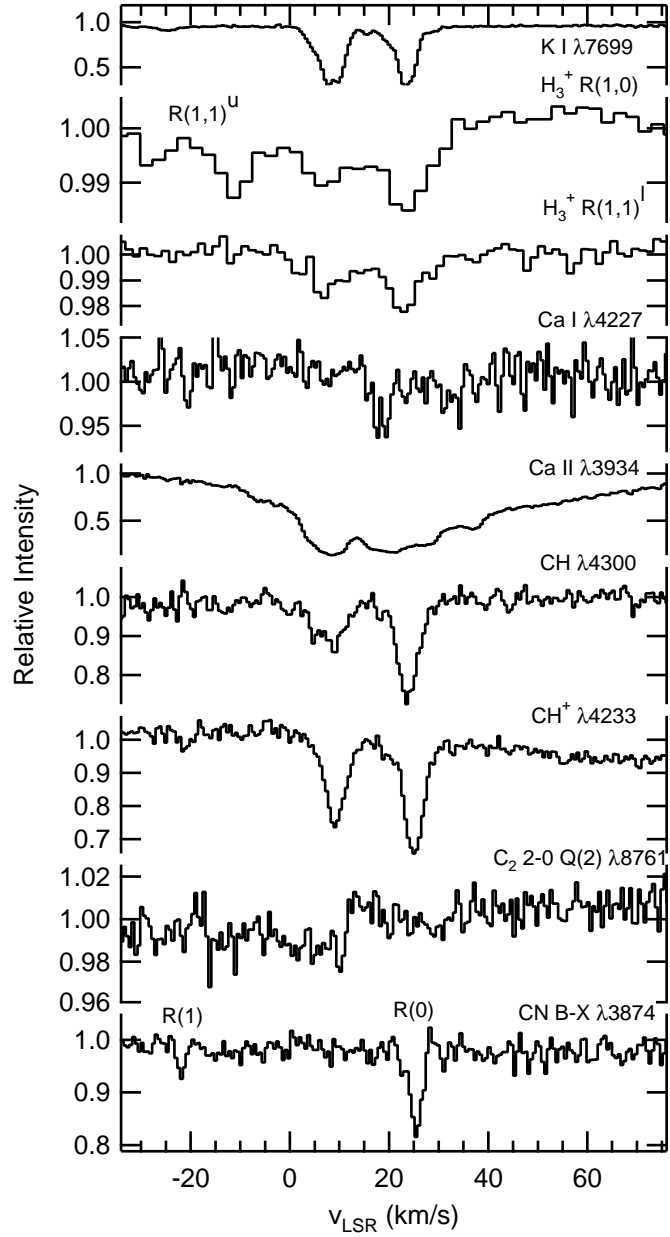}
\caption{Summary of spectra of HD 183143, in velocity space.
\label{HD183143_vel}}
\end{center}
\end{figure}

\begin{figure}
\begin{center}
\epsscale{0.54}
\plotone{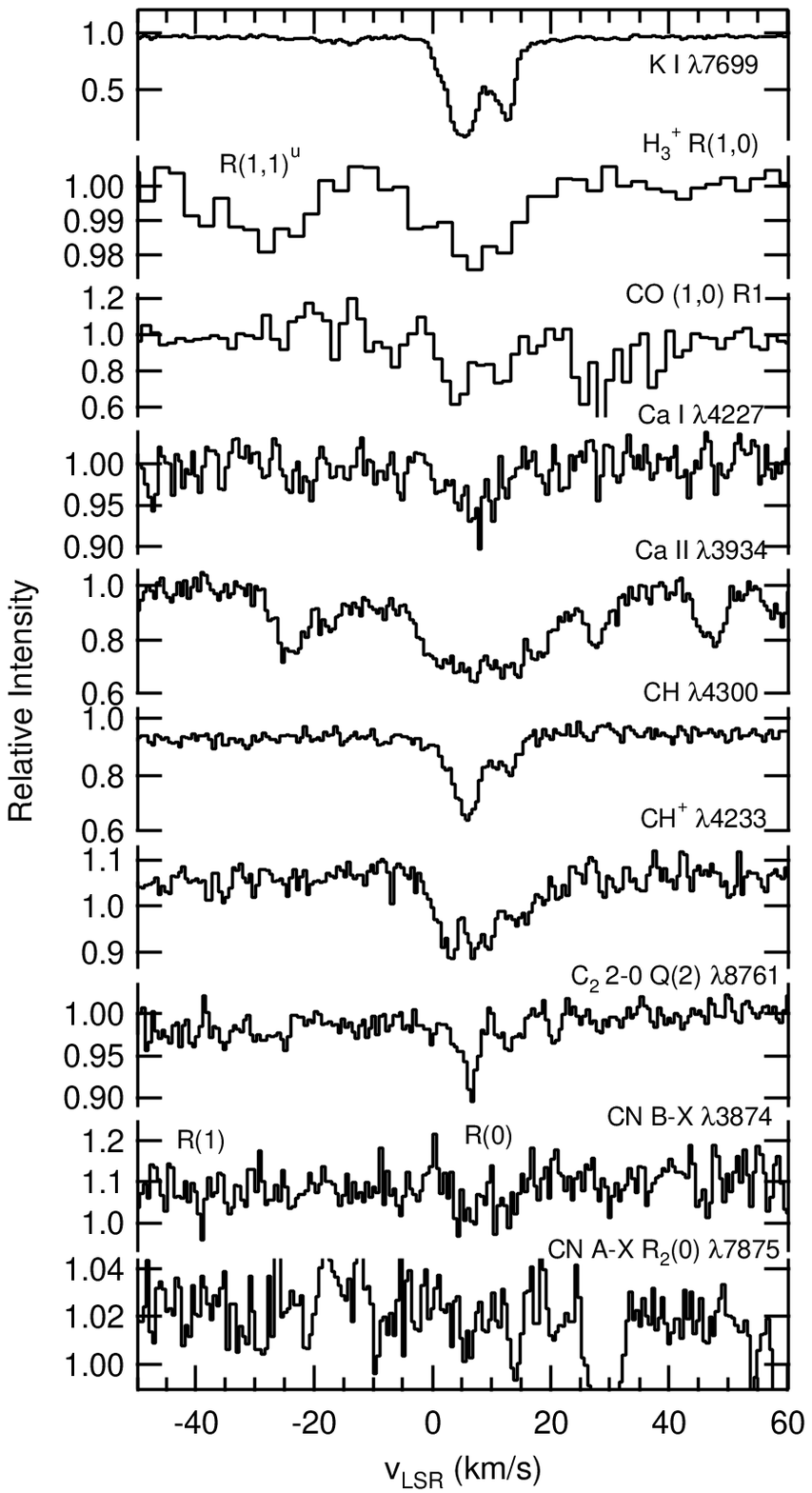}
\caption{Summary of spectra of Cyg OB2 5, in velocity space.  
The structure
near 30 and 60 km/s in the lower trace is due to atmospheric lines.
\label{CygOB25_vel}}
\end{center}
\end{figure}

\begin{figure}
\begin{center}
\epsscale{0.54}
\plotone{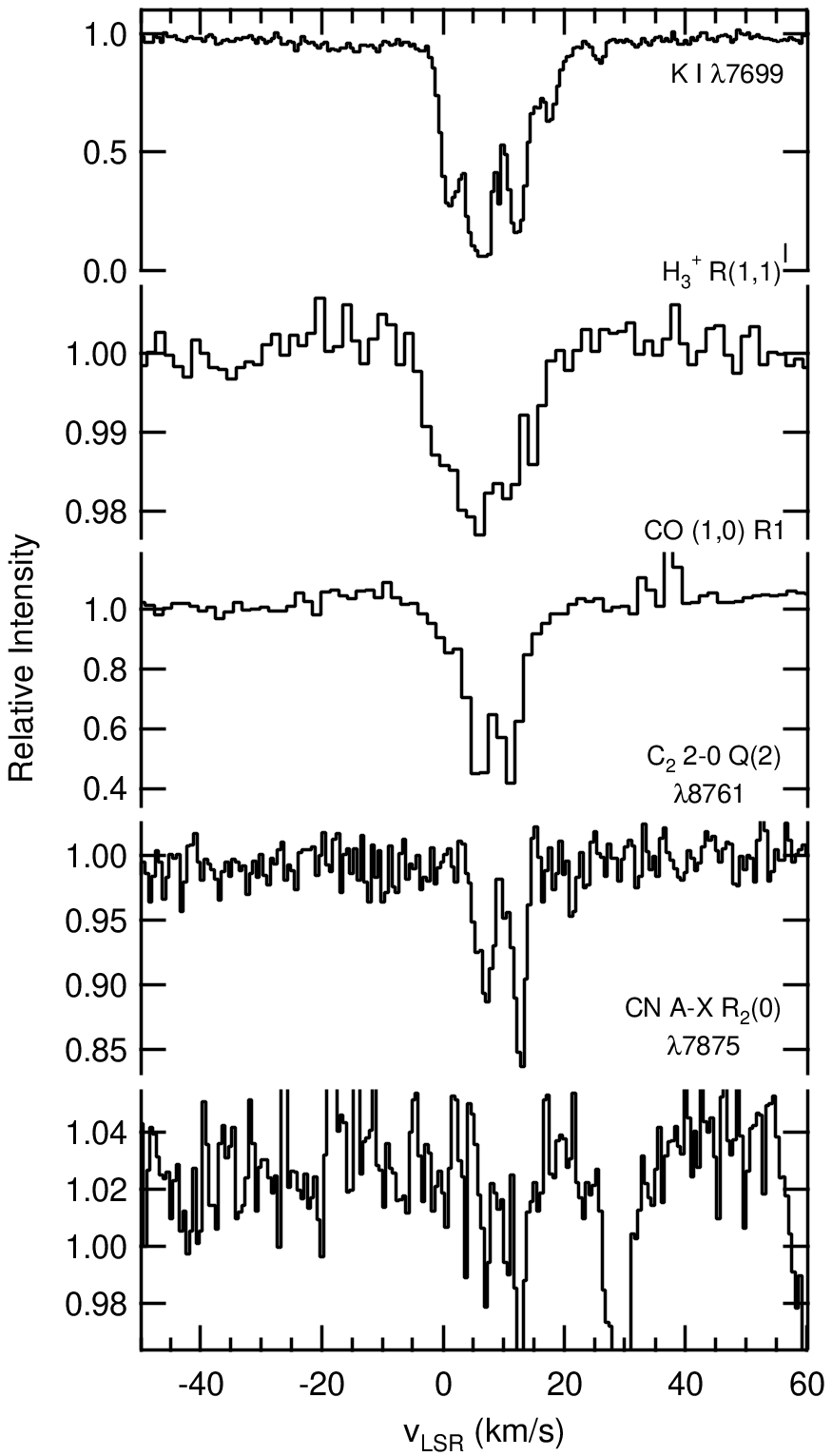}
\caption{Summary of spectra of Cyg OB2 12, in velocity space.
The structure
near 30 and 60 km/s in the lower trace is due to atmospheric lines.
\label{CygOB212_vel}}
\end{center}
\end{figure}

The results of our high resolution visible spectroscopy are shown in
Figure \ref{HD183143_vel} (HD 183143), Figure \ref{CygOB25_vel} (Cygnus OB2 5),
and Figure \ref{CygOB212_vel} (Cygnus OB2 12).  These figures are plotted in
velocity (with respect to the local standard of rest), and also display the
infrared measurements of \hhh\ and CO.

CH appears as a doublet in Cygnus OB2 5, with velocities consistent with the
infrared CO measurements.  CH also appears as a doublet in HD 183143, with
remarkably similar velocity structure to the \hhh.  The blue wavelength
(4300 \AA) of the CH transition made it inaccessible in the case of Cygnus
OB 12 (which is very heavily reddened), but it may be possible to detect the
line with a larger telescope such as the Hobby-Eberly Telescope.  The line
parameters of CH (as well as CH$^+$ and CN) are listed in Table \ref{vistable}.
The CH column densities have been derived using the curve
of growth of \citet{EvD:CH}, assuming $b$=1 km/s.

It is interesting to compare the observed CH column densities with those
which would be predicted from the empirical relation between H$_2$ and CH
\citep{magnani}:
\[
N({\rm H}_2) = 2.1 \times 10^7 N({\rm CH}) + 2.2 \times 10^{20} {\rm cm}^{-2}
\]
For HD 183143, E(B-V)=1.2, so N$_H \sim 7 \times 10^{21}$
[assuming the standard gas-to-dust conversion factor \citep{bohlin}], and
(if $f \equiv$ 2N(H$_2$)/[N(H)+2N(H$_2$)] = 2/3) 
N(H$_2$) $\sim 2.3 \times 10^{21}$ cm$^{-2}$.  From the empirical relation,
one would then predict N(CH) $\sim 1.0 \times 10^{14}$, roughly twice the
observed value.  For Cygnus OB2 5, with E(B-V) = 2.1, we estimate
N(H$_2$) $\sim 4 \times 10^{21}$ and therefore N(CH) $\sim 1.8 \times 10^{14}$
cm$^{-2}$, over three times the observed value.  Assuming our choice of
$b$-value is not too large, this suggests that these
two sources may be somewhat different chemically from the usual diffuse cloud
sources used to develop the empirical relation.  However, this discrepancy
may not indicate a drastic departure in chemical conditions, as other sources
with lower color excess have been observed with similar departures from the 
empirical relations \citep{danks}.  The CH discrepancy can be reduced by
adopting a lower value of $f$, but this would decrease $n$(\hhh), exacerbating
the problems discussed in \S \ref{inferred}.

\label{CH}

Another interesting molecule is CN: it is observed (marginally) as a doublet
in both Cygnus OB2 12 (A-X) and in Cygnus OB2 5 (A-X and B-X).  However,
only one component (at v$_{\rm LSR} \sim 25$ km/s) is observed toward
HD 183143 --- whereas \hhh, CH, and CH$^+$ all show two velocity components!  
The R(1) line is also detected toward HD 183143, and the population of the
$J=0$ and $J=1$ levels are consistent with the temperature of the cosmic
microwave background.

Finally, it is worth noting that, at least in the case of Cygnus OB2 12
(Figure \ref{CygOB212_vel}),
the velocity profile of \hhh\ appears more similar to that of K \textsc{i}
 than to C$_2$, 
CN, or CO.  This could be because these three molecules are concentrated in
the denser regions of the sightline, whereas \hhh\ (and K \textsc{i}) 
may exist over 
a larger path length.

\section{Discussion}

\subsection{Model of \hhh\ Chemistry in Diffuse Clouds}

The formation of \hhh, in dense or diffuse clouds, begins with the (cosmic ray
induced) ionization of H$_2$ to form H$_2^+$.  The generally assumed rate of
ionization is $\zeta \sim 3 \times 10^{-17}$ s$^{-1}$, so that the average
H$_2$ gets ionized roughly once every 10$^9$ years.  Once ionized, the
H$_2^+$ quickly reacts with another H$_2$ to form \hhh\ and an H atom.
This ion-neutral reaction proceeds with the Langevin rate constant of 
$2 \times 10^{-9}$ cm$^3$ s$^{-1}$, so that in a medium of H$_2$ density
100 cm$^{-3}$, the average H$_2^+$ must wait about 2 months to react to form
\hhh.  Clearly the initial ionization is the rate-limiting step, so we can
say that \hhh\ is formed at a rate of $\zeta$ n(H$_2$).

The destruction of \hhh\ is very different in diffuse clouds, compared to
dense clouds.  In dense clouds, the dominant destruction path is an ion-neutral
reaction with CO, with a rate constant of $\sim 2 \times 10^{-9}$ 
cm$^3$ s$^{-1}$.  In diffuse clouds, electrons are very abundant (due to
photoionization of C), so dissociative recombination dominates, with a rate
constant of $k_e = 4.6 \times 10^{-6}\ T_e^{-0.65}$ cm$^3$ s$^{-1}$
\citep{sundstrom}.  

Although the electrons produced by photoionization are probably formed at a 
high temperature (photons are available at energies up to 13.6 eV, and the
ionization potential of C is only 11.3 eV), they will be quickly
thermalized by collisions with H$_2$.  Such collisions will occur with a
Langevin rate constant (which scales as $\mu^{-1/2}$, where $\mu$ is the 
reduced mass) of $\sim 8 \times 10^{-8}$ cm$^3$ s$^{-1}$; in a medium
with n(H$_2$) = 100 cm$^{-3}$, an electron experiences such a collision about
once per day.  The lifetime of the electron is limited by radiative
recombination with C$^+$ ions, which occurs with a rate constant of
$\sim 2 \times 10^{-11}$ cm$^3$ s$^{-1}$ \citep{aldrovandi}
at the low temperatures ($\sim$30 K)
of diffuse clouds.  If n(C$^+$) = 10$^{-2}$ cm$^{-3}$, the average electron
lifetime is about 10$^5$ years.  Evidently the electrons will be very quickly
thermalized to the kinetic temperatures of the clouds.  The \hhh\ electron
recombination rate constant
is then $k_e \sim 5 \times 10^{-7}$ cm$^3$ s$^{-1}$, and
the destruction rate of \hhh\ can be given as $k_e$ $n$(\hhh) $n$($e$).

To calculate the number density of \hhh\ \citep{mccalldiffuse,geballediffuse}, 
we make the steady-state approximation
that the rates of formation and destruction are equal.  Thus,
\[
\zeta n({\rm H}_2) = k_e n({\rm H}_3^+) n(e)
\]
which can be rearranged to give
\[
n({\rm H}_3^+) = \frac{\zeta}{k_e} \frac{n({\rm H}_2)}{n(e)}
\]

The chemistry of \hhh\ is unique in the sense that the number density
of \hhh\ is not dependent upon the number density of the cloud, but only on
the ratio of the number densities of molecular hydrogen and electrons (as
well as the ratio of two constants $\zeta/k_e$).  We can further simplify
the above equation by considering $f$, the fraction of protons in H$_2$,
writing $n({\rm H}_2) = (f/2) [2n({\rm H}_2) + n({\rm H})] =
(f/2) n(\Sigma {\rm H})$.  Since most electrons are formed from the
ionization of carbon, we can also write $n(e) = (1-\alpha) n(\Sigma{\rm C})$,
where $\alpha$ is the fraction of carbon in un-ionized (neutral or molecular)
form.  Finally, the gas-phase ratio of carbon atoms to hydrogen atoms is
usually defined as $z_{\rm C} \equiv n(\Sigma {\rm C})/n(\Sigma {\rm H})$.
Putting this all together, we find
\begin{equation}
n({\rm H}_3^+) = \frac{\zeta}{k_e}\ \frac{f}{2}\ \frac{1}{1-\alpha}\ 
\frac{1}{z_{\rm C}}
\end{equation}

Adopting the values $\zeta = 3 \times 10^{-17}$ s$^{-1}$, 
$k_e = 5 \times 10^{-7}$ cm$^3$ s$^{-1}$, $f = 2/3$, $\alpha \sim 0$,
and $z_{\rm C} = 1.4 \times 10^{-4}$ \citep{cardelli,sofia}, we obtain
\begin{equation}
n({\rm H}_3^+) = 1.4 \times 10^{-7}\ {\rm cm}^{-3}
\end{equation}

The validity of the steady-state approximation can be checked by comparing
the timescale needed to reach steady-state with other relevant timescales.
The timescale needed to achieve steady-state is approximately the steady
state number density of \hhh\ divided by the formation rate, or
$n({\rm H}_3^+)/[\zeta n({\rm H}_2)]$, which for n(H$_2$) = 100
cm$^{-3}$ is about one year --- clearly much shorter than other relevant
timescales!

\subsection{Inferred Cloud Parameters}
\label{inferred}

One cloud parameter can be determined from the \hhh\ observations
independent of the chemical model --- the kinetic temperature of the gas.
This is possible because the {\em ortho} ($J$=1, $K$=0) and {\em para}
($J$=1, $K$=1) levels of \hhh\ are efficiently thermalized together 
through proton
hop and proton exchange reactions with H$_2$ \citep{mccallfaraday}.  These
reactions occur with the Langevin rate constant of $2 \times 10^{-9}$
cm$^3$ s$^{-1}$, so that the average \hhh\ experiences such a reaction
about every two months (assuming $n$(H$_2$) = 100 cm$^{-3}$).  The lifetime
of the average \hhh\ can be estimated from the dissociative recombination
rate $k_e$ and the number density of electrons $n(e) \sim 10^{-2}$ cm$^{-3}$
to be about 4.5 years, considerably longer than the thermalization timescale.
Therefore, we can use the Boltzmann expression
\[
\frac{N_{ortho}}{N_{para}} = \frac{g_{ortho}}{g_{para}} e^{-\Delta E /kT} = 2 e^{-32.87/T}
\]
to estimate the kinetic temperature from the observed {\em ortho:para} ratio
(in this equation, the $g$ values are the statistical weights).
For Cygnus OB2 12, we obtain T=27$\pm$4 K \citep{mccalldiffuse}; for 
Cygnus OB2 5, 47$\pm$13 K; for HD 183143, 31$\pm$3 K; for WR 104, 38$\pm$10 K; 
and for the +48 km/s component of WR 118, 40$\pm$3 K.  
[We cannot estimate the {\em ortho:para} ratio for the
low velocity component toward WR 118 due to the overlapping of the lines.]

Given our calculated number density of \hhh\ from the chemical model, 
we can now estimate the path length of absorption, using the relation
$L = N({\rm H}_3^+) / n({\rm H}_3^+)$.  Once the path length has been
calculated, we can then estimate the average number density 
of collision partners along the
path length as $\langle n \rangle = [ N({\rm H}_2) + N({\rm H}) ] / L$,
where $N$(H$_2$) and $N$(H) can be estimated from the color excess and
an assumed value of $f$ (2/3).  The results of this analysis are given in
Table \ref{parameters}.

In most cases, the derived pathlengths are a substantial fraction of
the estimated distance to the star, which seems difficult to accept.
In addition, the derived average number densities 
(for the sources in which \hhh\ was detected)
are in the range 1--5 cm$^{-3}$, which seems unreasonably low.  These
densities are nearly two orders lower than the densities typically
derived from the rotational excitation of C$_2$.  In addition, these
densities are so low that a substantial fraction of H$_2$ should be
photodissociated, meaning that our value of $f=2/3$ should be lowered,
which would in turn make the \hhh\ number density even smaller, the path
lengths even longer, and the average density even lower.

It seems clear that there is a serious problem with the model, and that
$n$(\hhh) must be larger (probably by at least one order of magnitude)
than we have calculated.  In the next subsection,
we explore possible solutions to this problem.

\subsection{Possible Solutions}

The equation for the number density of \hhh\
\begin{equation}
n({\rm H}_3^+) = \frac{\zeta}{k_e}\ \frac{f}{2}\ \frac{1}{1-\alpha}\ 
\frac{1}{z_{\rm C}}
\end{equation}
contains five parameters: the H$_2$ ionization rate ($\zeta$), the dissociative
recombination rate ($k_e$), the fraction of protons in molecular form ($f$),
the fraction of carbon atoms that are not ionized ($\alpha$), and the gas
phase carbon fraction ($z_{\rm C}$).  We now consider each one of these
parameters in more detail.

The fraction of protons in molecular form ($f$) has been assumed to have the
value 2/3, which is the largest value found in the studies of molecular and
atomic hydrogen by Copernicus, IUE, and FUSE.  The maximum value of $f$ is 1,
so even making this adjustment (as is done in the model of
\citet{cp:dalgarno}) will only increase $n$(\hhh) by a factor of
1.5 --- hardly enough to fix the problem.  In addition, increasing the value
of $f$ would increase the discrepancy between inferred and measured $N$(CH),
as discussed in \S \ref{CH}.

The gas phase carbon fraction ($z_{\rm C}$) has been taken to be 
$1.4 \times 10^{-4}$ based on ultraviolet observations of classical diffuse
clouds \citep{cardelli,sofia}.  Lowering this value (for instance, by assuming
that a larger fraction of the carbon is depleted onto grains) would 
reduce the number density of electrons and therefore increase
the number density of \hhh.  However, there is no astronomical evidence to
support the magnitude of depletion necessary to mitigate the \hhh\ problem.

The fraction of carbon that is not ionized ($\alpha$) has been assumed to be
near zero.  It is conceivable that a substantial fraction of the
carbon could be in the form of neutral C atoms (rather than C$^+$), and there
is no observational evidence (in these particular sightlines) that requires
most of the carbon to be ionized.  However, most chemical models of translucent
clouds (see, for example, Figure 9 of \citet{EvD88})
suggest that when the optical depth has increased to the point where
$n$(C)=$n$(C$^+$), over 10\% of the carbon atoms
are already in the form of CO.  Taken along with our observations of a low
column density of CO in these sightlines, this suggests that $\alpha$ must
be less than 0.5, and thus could contribute at most a factor of 2 towards
increasing the \hhh\ number density.  

However, other models (e.g.\ \citet{federman}) suggest that the abundance 
of C$^+$ in an $A_{\rm V}=4$ cloud may be 15 times lower than in 
an $A_{\rm V}=1$ cloud.  If we adopt this model, then it is possible that
$\alpha \stackrel{>}{_{\sim}} 0.9$, which would increase $n$(\hhh) by one
order of magnitude.  Given the differences between the models, we hesitate 
to rely on them too much, in the absence of observational evidence.  The best
determination of the value of $\alpha$ would come from an observation of the
C$^+$ transitions at 1334 and 2325 \AA\ using the Hubble Space Telescope.

Along the same lines, one might speculate about the possibility of an
``electron sink'' in these sightlines, so that C could still be ionized but
the free electron abundance would be low.  The best candidate mechanism for
removing electrons from the gas is probably attachment to grains or large
molecules (e.g., \citet{lepp:dalgarno}).  However, because the number density
of large molecules or grains is orders of magnitude lower than the number
density of electrons (since these molecules or grains form from elements with
cosmic abundance less than or equal to that of carbon, and each molecule or
grain contains many atoms), this process could
not effect the removal of a significant fraction of the electrons from the
gas phase.

The H$_2$ ionization rate ($\zeta$) has been assumed to be $3 \times 10^{-17}$
s$^{-1}$, but this value is not terribly well constrained.  
\citet{cp:dalgarno} assume $6 \times 10^{-17}$ in order to
increase the \hhh\ number density, and values as high as $\sim 2 \times
10^{-16}$ have been derived from analysis of the chemistry leading to OH
in diffuse clouds \citep{EvD:black}.  While the flux of high energy 
($\stackrel{>}{_{\sim}}$ 100 MeV) cosmic rays can be constrained by
observations in the interplanetary medium, the flux of lower energy cosmic 
rays is essentially unconstrained, due to the influence of the Sun.
If the cosmic ray spectrum is assumed not to roll off too rapidly
below 100 MeV,
it is conceivable that $\zeta$ might be as high as $10^{-15}$ s$^{-1}$
\citep{hayakawa}.  A large flux of low energy cosmic rays might increase the
\hhh\ number density in diffuse clouds while not seriously affecting the
situation in dense clouds, into which the low energy cosmic rays could not
penetrate.

Other sources of H$_2$ ionization have also been suggested.  
\citet{black:rs} has suggested that X-rays from the luminous stars in the
Cygnus OB2 association could increase the effective $\zeta$ and therefore
$n$(\hhh).  While this suggestion might solve the problem for the sightlines
toward Cygnus OB2, our observations of high $N$(\hhh) toward several other
sources
imply that a more general solution is needed.  \citet{black:rs}
has also suggested that ultraviolet photoionization of H$_2$ might contribute.
However, photoionization of H$_2$ requires photons with energies above 15.4 eV,
which will ionize H atoms.  While Black suggests that such photons could
escape the H \textsc{ii} 
region near the Cygnus OB2 giants, it seems unlikely that
they could penetrate the boundaries of diffuse clouds, where H atoms are
abundant.  Furthermore, our new detections of \hhh\ again suggest that 
the sightlines toward the Cygnus OB2 association are not unique.

The last parameter in the equation for the \hhh\ number density is the
dissociative recombination rate ($k_e$).  In this work, we have adopted the
value of $k_e = 4.6 \times 10^{-6}\ T_e^{-0.65}$ cm$^3$ s$^{-1}$ derived
from storage ring experiments \citep{sundstrom}.  However, the
value of this constant has been the matter of great controversy over the
past two decades --- other experimental techniques currently 
yield a value of $k_e$
that is about one order of magnitude lower (for a review of the field, see
\citet{larsson:rs}).  It has been suggested that the discrepancy might be due
to stray fields present in the storage ring experiments, and that the
recombination rate under interstellar conditions could be quite low.  To
make matters worse, attempts to theoretically calculate the recombination
rate yield rates more than an order of magnitude smaller than the smallest
values obtained in experiments \citep{orel:rs}, but it should be noted
that the theory of this recombination process is not yet mature.
Given the present uncertainty in the true value of $k_e$, it is
possible that the problem of diffuse cloud \hhh\ may be solved on this front.

\subsection{Observational Tests}

\begin{figure}
\begin{center}
\epsscale{1}
\plotone{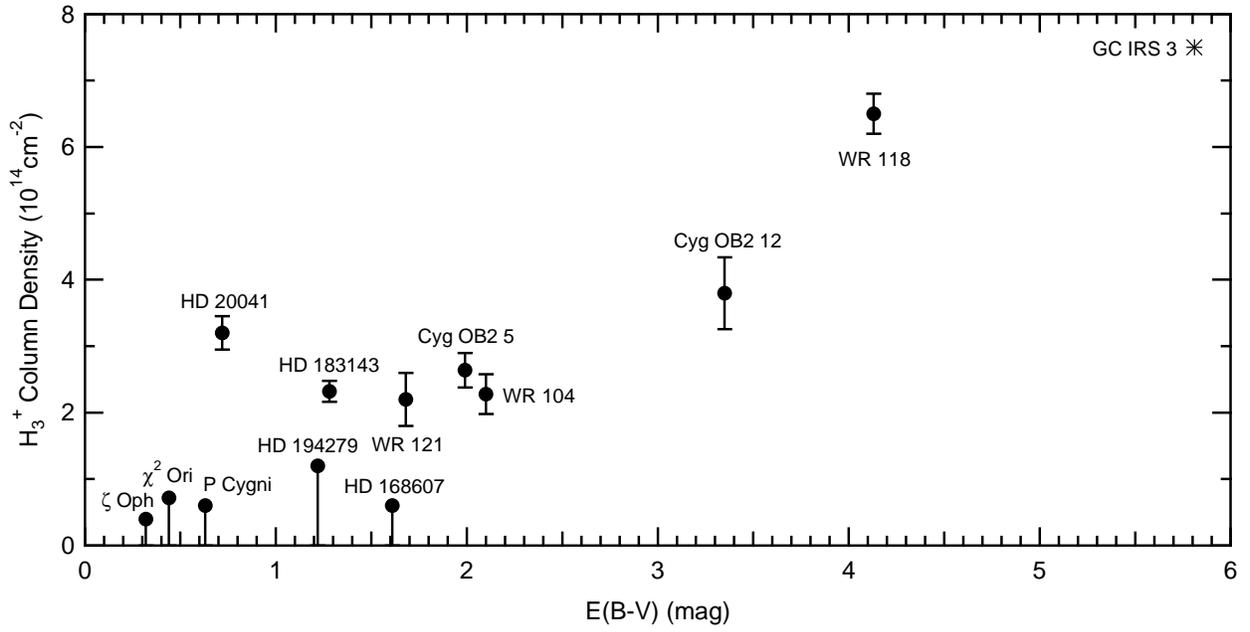}
\caption{\hhh\ Column Density versus color excess.
\label{EB_V}}
\end{center}
\end{figure}

The resolution of the mystery of diffuse cloud \hhh\ will most likely come
from further observational work.  First we must determine with certainty
whether the unexpectedly high number density of \hhh\ is a ``local'' problem
(i.e., due to special conditions in these particular sightlines, or this
class of sightlines), or if it is a ``global'' problem.  The present
observational situation is summarized in Figure \ref{EB_V}, which plots
the observed \hhh\ column density (or the upper limit, denoted by a bar
to zero) versus the color excess E(B-V).  In addition to the present 
observations, the detection of \hhh\ toward the Galactic Center source
GC IRS 3 \citep{geballediffuse} has been added (the \hhh\ column density plotted
is that of the narrow component attributed to diffuse clouds by
\citep{geballediffuse}, and the adopted color excess is that
attributed to diffuse clouds by \citep{whittet}).
Keeping in mind that 
\begin{equation}
N({\rm H}_3^+) \sim n({\rm H}_3^+)\ L \propto n({\rm H}_3^+)\ \frac{E(B-V)}{\langle n_{\rm H} \rangle}
\end{equation}
the plot is not inconsistent with a constant $n$(\hhh) and small variations
(of a factor of a few) in the average density $\langle n_{\rm H} \rangle$.
Therefore at this stage there is no strong evidence that the ``local''
parameters ($f$, $\alpha$, and $z_{\rm C}$) are any different in the detection
sources than they are in the non-detections such as $\zeta$ Oph.

However, this inference needs to be tested directly by observations.  This
can be achieved by detecting \hhh\ along a line of sight with lower
E(B-V) which can be studied by FUSE and HST.
Once the column densities
of \hhh, H \textsc{i}, H$_2$, C \textsc{i}, C \textsc{ii}, 
CO, and CH are obtained for a single sightline,
the influence of the ``local'' parameters will be directly determined.

Assuming that these parameters are found to have their canonical values, we
will be left with the ratio $\zeta / k_e$.
Absent a speedy resolution
from the experimenters or theorists in the field of dissociative recombination,
further insight into these two constants
 can perhaps best be obtained through
an observational survey of \hhh\ in more heavily reddened lines of sight.
The rate of \hhh\ destruction is controlled by $n(e)/n_{\rm H}$
(which depends on the optical depth in the ultraviolet), whereas the rate of
\hhh\ formation is controlled by $\zeta$ (which, with a given incident
cosmic ray spectrum, depends on the stopping power of the cloud as a function
of cosmic ray energy).  Therefore, a detailed study of the diffuse-to-dense
cloud transition, along with chemical models, could help constrain the values 
of $\zeta$ and $k_e$.  The highly-reddened, early-type subset of the Stephenson
catalog recently compiled by \citet{rawlings}
may serve as a good starting point for such a study.

\section{Conclusions}

The \hhh\ molecular ion has now been definitely detected in seven
diffuse cloud lines of sight, suggesting that
its unexpectedly high abundance is not due to the peculiarities of a
particular region (the Cygnus OB2 association), but rather a general
feature of the diffuse interstellar medium.  \hhh\ is observed in clouds
with and without CO, C$_2$, and CN, confirming that the chemistry that leads
to \hhh\ is completely decoupled from that which is responsible for these
heavier diatomics.  The most likely explanation for the high \hhh\ abundance
is a larger than expected $\zeta / k_e$ ratio --- due either to a larger
flux of low-energy cosmic rays or to a lower value of $k_e$ in interstellar
conditions.  The possibility of a lower than expected electron density has 
not been ruled out, and should be directly tested by
finding \hhh\ in a less reddened source.

\acknowledgements

The authors wish to acknowledge helpful discussions with D.~G.\ York and
L.~M.\ Hobbs.  We are grateful to the staffs of the various observatories
we have used, as well as to the respective telescope allocation committees.
This work has made use of the NASA Astrophysics Data Service, as well as
the SIMBAD database at the Centre de Donn\'{e}es astronomiques de Strasbourg.
N.J.E.\ acknowledges support from the State of Texas and NSF grant
9988230.
The University of Chicago portion of this work has been supported by NSF
grant PHYS-9722691 and NASA grant NAG5-4070.  B.J.M.\ has been supported by
the Fannie and John Hertz Foundation, and wishes to acknowledge a
Sigma Xi Grant-in-Aid of Research and NOAO for travel support.




\clearpage


\clearpage

\clearpage

\clearpage

\clearpage

\clearpage

\clearpage

\clearpage

\clearpage

\clearpage

\clearpage

\clearpage

\clearpage

\clearpage

\clearpage

\clearpage

\begin{table}

\begin{center}
\caption{Stellar Parameters.\label{targets}}

{

\begin{tabular}{lcccc}
Name         & V    & E(B-V) & Spectral Type & Distance (pc)
\\ \tableline
WR 118       & 22.0 & 4.13\tablenotemark{a} & WC9    & 6300\tablenotemark{b} \\
Cyg OB2 12   & 11.4 & 3.35\tablenotemark{c} & B5Iab  & 1700\tablenotemark{d} \\
WR 104       & 13.5 & 2.10\tablenotemark{a} & WC9    & 1300\tablenotemark{b} \\
Cyg OB2 5    & 9.2  & 1.99\tablenotemark{c} & O7e    & 1700\tablenotemark{d} \\
WR 121       & 11.9 & 1.68\tablenotemark{a} & WC9    & 1690\tablenotemark{b} \\
HD 168607    & 8.25 & 1.61\tablenotemark{e} & B9Ia   & 1100\tablenotemark{f} \\
HD 183143    & 6.9  & 1.28\tablenotemark{e} & B7Ia   & 1000\tablenotemark{f} \\
HD 194279    & 7.1  & 1.22\tablenotemark{e} & B1.5Ia & 1100\tablenotemark{f} \\
HD 20041     & 5.8  & 0.70\tablenotemark{g} & A0Ia   & 1400\tablenotemark{f} \\
P Cygni      & 4.8  & 0.63\tablenotemark{h} & B2pe   & 1800\tablenotemark{h} \\
$\chi^2$ Ori & 4.6  & 0.44\tablenotemark{i} & B2Iae  & 1000\tablenotemark{f} \\
$\zeta$ Oph  & 2.6  & 0.32\tablenotemark{e} & O9V    & 140\tablenotemark{j} \\

\end{tabular}
}

\tablenotetext{a}{From \citet{pendleton}, assuming 
$R_V$ = $A_V$ / E(B-V) = 3.1.}

\tablenotetext{b}{\citet{wolfrayet}.}

\tablenotetext{c}{From \citet{schulte}.}

\tablenotetext{d}{\citet{cygdist}.}

\tablenotetext{e}{From \citet{snowyorkwelty}.}

\tablenotetext{f}{Our estimate (spectroscopic parallax).}

\tablenotetext{g}{\citet{racine}.}

\tablenotetext{h}{From \citet{pcygni}.}

\tablenotetext{i}{From the intrinsic color of \citet{wegner}.}

\tablenotetext{j}{From Hipparcos catalog \citep{hipparcos}.}

\end{center}
\end{table}

\clearpage

\begin{table}

\begin{center}
\caption{Log of observations.\label{obslog}}

{
\scriptsize
\begin{tabular}{lcccccc}
Date (UT) & Observatory & Species & Transition & Object & Standard & Time (min)
\\ \tableline
Jun 25, 1997 & KPNO & CO & $v$=2-0 & HD 183143 & $\alpha$ Lyr & 60 \\
Jul 3, 1998 & KPNO & \hhh & $R$(1,1)$^l$ & Cyg OB2 12 & $\alpha$ Cyg & 30 \\
Jul 3, 1998 & KPNO & \hhh & $R$(1,1)$^l$ & Cyg OB2 5 & $\alpha$ Cyg & 164 \\
Jul 4, 1998 & KPNO & \hhh & $R$(1,1)$^l$ & WR 104 & $\alpha$ Lyr & 35 \\
Jul 4, 1998 & KPNO & \hhh & $R$(1,1)$^l$ & WR 121 & $\alpha$ Lyr & 112 \\
Jul 4, 1998 & KPNO & \hhh & $R$(1,1)$^l$ & WR 118 & $\alpha$ Cyg & 46 \\
Jul 4, 1998 & KPNO & \hhh & $R$(1,1)$^l$ & HD 194279 & $\alpha$ Cyg & 92 \\
Jul 4, 1999 & KPNO & CO & $v$=2-0 & Cyg OB2 12 & $\alpha$ Cyg & 4 \\
Jul 5, 1999 & KPNO & CO & $v$=1-0 & Cyg OB2 12 & $\alpha$ Lyr & 4 \\
Jul 5, 1999 & KPNO & CO & $v$=1-0 & Cyg OB2 5 & $\alpha$ Lyr & 8 \\
Jun 25, 2000 & KPNO & \hhh & $R$(1,1)$^l$ & Cyg OB2 12 & $\alpha$ Cyg & 60 \\
Jun 25, 2000 & KPNO & \hhh & $R$(1,1)$^l$ & HD 183143 & $\alpha$ Lyr & 84 \\
Jun 25, 2000 & KPNO & \hhh & $R$(1,1)$^l$ & P Cygni & $\alpha$ Lyr & 78 \\
Jun 26, 2000 & KPNO & \hhh & $R$(1,1)$^l$ & $\zeta$ Oph & $\alpha$ Lyr & 48 \\
Jun 26, 2000 & KPNO & CO & $v$=1-0 & Cyg OB2 12 & $\alpha$ Lyr & 36 \\
Jun 26, 2000 & KPNO & CO & $v$=1-0 & Cyg OB2 5 & $\alpha$ Lyr & 90 \\
Mar 13, 2001 & KPNO & \hhh & $R$(1,1)$^l$ & HD 20041 & $\alpha$ CMa & 100 \\
Mar 13, 2001 & KPNO & \hhh & $R$(1,1)$^l$ & $\chi^2$ Ori & $\alpha$ CMa & 60 \\
Jul 6, 2000 & UKIRT & \hhh & $R$(1,0),$R$(1,1)$^u$ & HD 183143 & $\alpha$ Lyr &
29 \\
Jul 6, 2000 & UKIRT & \hhh & $R$(1,0),$R$(1,1)$^u$ & Cyg OB2 5 & $\alpha$ Cyg &
43 \\
Jul 7, 2000 & UKIRT & \hhh & $R$(1,0),$R$(1,1)$^u$ & HD 168607 & $\pi$ Sgr & 25
\\
May 24, 2001 & UKIRT & \hhh & $R$(1,0),$R$(1,1)$^u$ & WR 104 & $\eta$ Oph & 8 \\
May 24, 2001 & UKIRT & \hhh & $R$(1,0),$R$(1,1)$^u$ & WR 118 & $\alpha$ Cyg & 6
\\
May 28, 2001 & UKIRT & CO & $v$=1-0 & HD 183143 & $\alpha$ Aql & 24 \\
May 28, 2001 & UKIRT & CO & $v$=1-0 & WR 104 & $\beta^1$ Sco & 3 \\
May 28, 2001 & UKIRT & CO & $v$=1-0 & WR 118 & $\eta$ Oph & 5 \\
Sep 12, 2000 & McDonald &
\( \left\{ \begin{array}{l} {\rm K\ I} \\ {\rm CN} \end{array} \right. \) &
\( \left. \begin{array}{c} ^2P - {^2S} \\ \mbox{{\rm A-X}}\ v \mbox{\rm =2-0} 
\end{array} \right\} \)  &
Cyg OB2 12 & $\alpha$ Cyg & 60 \\
Sep 12, 2000 & McDonald & '' & '' & Cyg OB2 5 & $\alpha$ Cyg & 60 \\
Sep 12, 2000 & McDonald & '' & '' & HD 183143 & $\alpha$ Cyg & 60 \\
Sep 12, 2000 & McDonald & C$_2$ & A-X $v$=2-0 & Cyg OB2 12 & $\alpha$ Cep & 60 \\
Sep 12, 2000 & McDonald & C$_2$ & A-X $v$=2-0 & Cyg OB2 5 & $\alpha$ Cep & 60 \\
Sep 12, 2000 & McDonald & C$_2$ & A-X $v$=2-0 & HD 183143 & $\alpha$ Cep & 60 \\
Sep 14, 2000 & McDonald &
\( \left\{ \begin{array}{l} {\rm Ca\ I} \\ {\rm Ca\ II} \\ {\rm CH} \\ {\rm CH^+} 
\\ {\rm CN} \end{array} \right. \) &
\( \left. \begin{array}{c} ^1P - {^1S} \\ ^2P - {^2S} \\
\mbox{{\rm A-X}}\ v \mbox{\rm =0-0} \\
\mbox{{\rm A-X}}\ v \mbox{\rm =0-0} \\
\mbox{{\rm B-X}}\ v \mbox{\rm =0-0}
\end{array} \right\} \)  &
Cyg OB2 5 & $\alpha$ Cyg & 200 \\
Sep 14, 2000 & McDonald & '' & '' & HD 183143 & $\alpha$ Cyg & 100 \\
Jul 28, 1999 & CSO & $^{12}$CO & $J$=2-1 & Cyg OB2 12 & 30' S & 14 \\
Jul 28, 1999 & CSO & $^{12}$CO & $J$=2-1 & Cyg OB2 5 & 30' S & 14 \\
Jul 28-29, 1999 & CSO & $^{13}$CO & $J$=2-1 & Cyg OB2 12 & 30' S & 32 \\
Jul 28-29, 1999 & CSO & $^{13}$CO & $J$=2-1 & Cyg OB2 5 & 30' S & 32 \\
Jul 28, 1999 & CSO & $^{13}$CO & $J$=3-2 & Cyg OB2 12 & 30' S & 20 \\
Jul 28, 1999 & CSO & $^{13}$CO & $J$=3-2 & Cyg OB2 5 & 30' S & 20 \\
Apr 9, 2000 & NRO & $^{12}$CO & $J$=1-0 & Cyg OB2 12 & ($l$=81,$b$=3) & 18 \\
Apr 9, 2000 & NRO & $^{13}$CO & $J$=1-0 & Cyg OB2 12 & ($l$=81,$b$=3) & 37 \\
Apr 11, 2000 & NRO & $^{12}$CO & $J$=1-0 & Cyg OB2 ``A'' & ($l$=81,$b$=3) & 10 \\
Apr 11, 2000 & NRO & $^{12}$CO & $J$=1-0 & Cyg OB2 ``B'' & ($l$=81,$b$=3) & 10 \\
Apr 11, 2000 & NRO & $^{12}$CO & $J$=1-0 & Cyg OB2 ``C'' & ($l$=81,$b$=3) & 15 \\
Apr 11, 2000 & NRO & $^{12}$CO & $J$=1-0 & Cyg OB2 ``D'' & ($l$=81,$b$=3) & 10 \\
Apr 14, 2000 & NRO & $^{12}$CO & $J$=1-0 & Cyg OB2 5 & ($l$=81,$b$=3) & 13 \\
Oct 13, 2000 & JCMT & $^{12}$CO & $J$=2-1 & HD 183143 & (8 MHz) & 60 \\
Oct 13, 2000 & JCMT & $^{12}$CO & $J$=2-1 & HD 183143 & (16 MHz) & 60 \\
Dec 14, 2000 & NRO & $^{12}$CO & $J$=1-0 & HD 183143 & (3 MHz) & 12 \\
Dec 14, 2000 & NRO & $^{12}$CO & $J$=1-0 & HD 183143 & (6 MHz) & 13 \\
Dec 15, 2000 & NRO & $^{12}$CO & $J$=1-0 & HD 183143 & (4 MHz) & 31 \\
Dec 15, 2000 & NRO & $^{12}$CO & $J$=1-0 & HD 183143 & (9 MHz) & 45 \\
\end{tabular}
}
\end{center}
\end{table}

\clearpage

\begin{table}
{
\begin{center}
\footnotesize
\caption{\hhh\ Line Parameters.\label{h3pluslines}}

\begin{tabular}{clcccccc}
Object & \multicolumn{1}{c}{Line} & v$_{\rm LSR}$ & FWHM &
W$_{\lambda}$ & $\sigma$(W$_{\lambda}$)\tablenotemark{a} & N(\hhh) & 
$\sigma$(N)\tablenotemark{a} \\
 & & (km/s) & (km/s) & (\AA) & (\AA) & (10$^{14}$ cm$^{-2}$) & (10$^{14}$ cm$^{-2}$) \\
 \tableline
Cyg OB2 12 & $R$(1,1)$^l$ & 7.0 & 14.6 & 0.044 & 0.002 & 2.02 & 0.09 \\
Cyg OB2 5 & $R$(1,1)$^l$ & 7.3 & 11.5 & 0.022 & 0.003 & 0.99 & 0.15 \\
          & $R$(1,1)$^u$ & 5.6 & 15.2 & 0.035 & 0.004 & 1.43 & 0.16 \\
          & $R$(1,0)     & 5.8 & 15.4 & 0.048 & 0.004 & 1.21 & 0.10 \\
WR 121 & $R$(1,1)$^l$ & 9.7 & 17.8 & 0.024 & 0.004 & 1.12 & 0.20 \\
WR 104 & $R$(1,1)$^l$ & 10.9 & 15.1 & 0.028 & 0.002 & 1.27 & 0.07 \\
       & $R$(1,1)$^u$ & 11.2 & 14.6 & 0.030 & 0.005 & 1.23 & 0.21 \\
       & $R$(1,0)     & 7.9  & 18.5 & 0.041 & 0.006 & 1.05 & 0.14 \\
WR 118 & $R$(1,1)$^l$ & 4.8 & 9.5 & 0.019 & 0.001 & 0.89 & 0.06 \\
       &              & 47.6 & 12.3 & 0.041 & 0.002 & 1.88 & 0.07 \\
       & $R$(1,1)$^u$ & 3.7 & 17.1 & 0.045 & 0.003 & 1.86 & 0.11 \\
       &              & 44.3\tablenotemark{b} & 15.4 & 0.026 & 0.002 & 1.06 & 0.10 \\
       & $R$(1,0)     & 8.5\tablenotemark{b} & 21.7 & 0.077 & 0.003 & 1.95 & 0.07 \\
       &              & 48.1 & 19.6 & 0.065 & 0.003 & 1.64 & 0.07 \\
HD 183143 & $R$(1,1)$^l$ & 8.3\tablenotemark{c} & 10.5 & 0.020 & 0.002 & 0.94 & 0.08 \\
          &              & 23.6 & 8.4 & 0.024 & 0.002 & 1.10 & 0.07 \\
          & $R$(1,1)$^u$ & 7.1 & 12.2 & 0.012 & 0.001 & 0.50 & 0.05 \\
          &              & 23.4 & 9.9 & 0.017 & 0.001 & 0.70 & 0.05 \\
          & $R$(1,0) & 5.6 & 14.0 & 0.022 & 0.001 & 0.54 & 0.03 \\
          &          & 22.1 & 11.0 & 0.023 & 0.001 & 0.59 & 0.03 \\
HD 20041  & $R$(1,1)$^l$ & -0.5 & 11.5 & 0.038 & 0.002 & 1.74 & 0.09 \\
HD 194279 & $R$(1,1)$^l$ & --- & 10\tablenotemark{d} & --- & 0.005 & --- 
& 0.21\tablenotemark{e} \\
P Cygni & $R$(1,1)$^l$ & --- & 10\tablenotemark{d} & --- & 0.002 & --- 
& 0.10\tablenotemark{e} \\
$\zeta$ Oph & $R$(1,1)$^l$ & --- & 10\tablenotemark{d} & --- & 0.001 & --- 
& 0.06\tablenotemark{e} \\
HD 168607 & $R$(1,1)$^u$ & --- & 10\tablenotemark{d} & --- & 0.003 & --- 
& 0.13\tablenotemark{e} \\
          & $R$(1,0)     & --- & 10\tablenotemark{d} & --- & 0.003 & --- 
& 0.07\tablenotemark{e} \\
$\chi^2$ Ori & $R$(1,1)$^l$ & --- & 10\tablenotemark{d} & --- & 0.003 & --- 
& 0.12\tablenotemark{e}  \\
\end{tabular}

\mbox{}

\tablenotetext{a}{Statistical uncertainties (1$\sigma$) are listed.  
Systematic errors
(for instance, due to ratioing of telluric lines) are difficult to estimate
and are likely larger.}

\tablenotetext{b}{These two features are overlapped, so the fit results are highly uncertain.}

\tablenotetext{c}{Affected by telluric line.}

\tablenotetext{d}{Adopted FWHM for purposes of computing upper limits.}

\tablenotetext{e}{The firm upper limit for N(\hhh) should be considered to be 3$\sigma$.}

\end{center}
}

\end{table}

\clearpage

\begin{table}
{
\begin{center}
\footnotesize
\caption{Infrared $^{12}$CO $v$=1-0 Line Parameters.\label{ircolines}}

\begin{tabular}{clcccccc}
Object & \multicolumn{1}{c}{Line} & v$_{\rm LSR}$ & FWHM &
W$_{\lambda}$ & $\sigma$(W$_{\lambda}$)\tablenotemark{a} & 
N(thin)\tablenotemark{b} & $\sigma$(N)\tablenotemark{a} \\
 & & (km/s) & (km/s) & (\AA) & (\AA) & (10$^{15}$ cm$^{-2}$) & (10$^{15}$ cm$^{-2}$) \\
 \tableline
Cyg OB2 12 & $P2$ &  5.9 & 4.3 & 0.199 & 0.021 & 2.28 & 0.24 \\
           &      & 11.8 & 6.1 & 0.300 & 0.025 & 3.42 & 0.29 \\
           & $P1$ &  6.0 & 4.1 & 0.344 & 0.024 & 4.71 & 0.32 \\
           &     & 10.3\tablenotemark{c} & 1.8 & 0.127 & 0.015 & 1.74 & 0.21 \\
           & $R0$ &  7.5 & 6.0 & 0.598 & 0.029 & 2.75 & 0.13 \\
           &      & 13.1 & 3.8 & 0.285 & 0.023 & 1.31 & 0.11 \\
           & $R1$ &  6.6 & 4.6 & 0.437 & 0.021 & 3.01 & 0.15 \\
           &      & 12.0 & 3.9 & 0.355 & 0.019 & 2.45 & 0.13 \\
           & $R2$ &  7.6 & 5.1 & 0.248 & 0.010 & 1.91 & 0.08 \\
           &      & 13.1 & 4.9 & 0.185 & 0.010 & 1.42 & 0.08 \\
Cyg OB2 5  & $R1$\tablenotemark{d} & 4.9 & 4.8 & 0.303 & 0.041 & 2.09 & 0.29 \\
           &      & 12.2 & 4.5 & 0.204 & 0.041 & 1.40 & 0.28 \\
HD 183143  & $R0$ & 25.6 & 13.0 & 0.138 & 0.013 & 0.64 & 0.06 \\
         & $R1$\tablenotemark{d} & 24.4 & 16.1 & 0.133 & 0.010 & 0.91 & 0.07 \\
WR 104     & $R0$ & 21.4 & 12.9 & 0.520 & 0.024 & 2.39 & 0.11 \\
           & $R1$ & 21.2 & 12.4 & 0.588 & 0.019 & 4.05 & 0.13 \\
           & $R2$ & 21.5 & 7.2 & 0.247 & 0.014 & 1.89 & 0.11 \\
           & $R3$ & 20.6 & 5.1 & 0.143 & 0.011 & 1.16 & 0.09 \\
WR 118   & $R0$ & 10.5 & 18.2 & 1.289 & 0.020 & 5.93 & 0.09 \\
         &      & 45.0\tablenotemark{e} & 18.0 & 1.147 & 0.020 & 5.28 & 0.09 \\
         &      & 59.4\tablenotemark{e} & 15.6 & 0.198 & 0.018 & 0.91 & 0.08 \\
         &      & 85.0 & 14.0 & 0.107 & 0.017 & 0.49 & 0.08 \\
         & $R1$ &  9.2 & 16.6 & 0.962 & 0.015 & 6.63 & 0.10 \\
         &      & 41.6\tablenotemark{e} & 15.1 & 1.078 & 0.014 & 7.43 & 0.10 \\
         &      & 56.9\tablenotemark{e} & 10.5 & 0.204 & 0.012 & 1.41 & 0.08 \\
         &      & 81.6 & 15.1 & 0.211 & 0.015 & 1.45 & 0.10 \\
         & $R2$ &  9.1 & 12.8 & 0.248 & 0.013 & 1.90 & 0.10 \\
         &      & 41.7\tablenotemark{e} & 10.6 & 0.539 & 0.012 & 4.13 & 0.09 \\
         &      & 53.6\tablenotemark{e} & 13.2 & 0.111 & 0.013 & 0.85 & 0.10 \\
         &      & 82.9 & 11.1 & 0.144 & 0.012 & 1.10 & 0.09 \\
         & $R3$ & 36.4 & 19.1 & 0.219 & 0.010 & 1.77 & 0.08 \\
         &      & 82.3 & 12.5 & 0.097 & 0.008 & 0.78 & 0.06 \\
\end{tabular}

\mbox{}

\tablenotetext{a}{Statistical uncertainties (1$\sigma$) are listed.  
Systematic errors
(for instance, due to ratioing of telluric lines) are difficult to estimate
and may be larger.}

\tablenotetext{b}{Listed column densities are for the lower state of each
absorption line in the given velocity component.  These are direct calculations
from the corresponding value of W$_{\lambda}$, in the optically thin limit.
For a more detailed analysis of the CO column density towards Cygnus OB2 12, 
see section \ref{cocyg12}.}

\tablenotetext{c}{Bad fit due to interference of telluric line.}

\tablenotetext{d}{Other lines too marginal to fit.}

\tablenotetext{e}{Blend of two components; individual fit parameters uncertain.}

\end{center}
}

\end{table}

\clearpage

\begin{table}
{
\begin{center}

\caption{mm-wave CO toward Cygnus OB2 12.\label{mmcolines}}

\begin{tabular}{cccccc}
Species   & Line & v$_{\rm LSR}$ & HWHM   & T$_A^*$ (peak) & Area \\
          &      & (km/s)        & (km/s) & (K)            & (K km/s) \\
 \tableline
$^{13}$CO & $J$=1-0 & 6.93     & 1.30     & 0.29 & 0.52 \\
          &         & 12.58    & 1.32     & 0.41 & 0.58 \\
          & $J$=2-1 & 6.9      & 1.39     & 0.21 & 0.31 \\
          &         & 12.4     & 1.35     & 0.33 & 0.47 \\
$^{12}$CO & $J$=1-0 & 7.31     & 2.52     & 1.49 & 3.34 \\
          &         & 12.72    & 1.81     & 3.08 & 5.98 \\
          & $J$=2-1 & 7.48     & 1.92     & 0.92 & 1.72 \\
          &         & 12.48    & 2.04     & 1.74 & 3.66 \\
\end{tabular}

\mbox{}

\end{center}
}

\end{table}

\clearpage

\begin{deluxetable}{ccccccccccc}
\rotate
\tabletypesize{\scriptsize}
\tablecaption{C$_2$ Line Parameters (components labelled as 1 and 2) 
for Cygnus OB2 12.\label{C2table}}\tablewidth{0pt}
\tablehead{
\colhead{Line} & 
\colhead{W$^1_{\lambda}$ \tablenotemark{a}}   & 
\colhead{N($J$)$^1$}   &
\colhead{v$^1_{\rm LSR}$} &
\colhead{FWHM$^1$}  & 
\colhead{W$^2_{\lambda}$ \tablenotemark{a}} & 
\colhead{N($J$)$^2$} &
\colhead{v$^2_{\rm LSR}$}     & 
\colhead{FWHM$^2$}  &
\colhead{W$^{tot}_{\lambda}$ \tablenotemark{a}}   & 
\colhead{N($J$)$^{tot}$} \\
\colhead{} &
\colhead{(m\AA)} &
\colhead{($10^{12}$ cm$^{-2}$)} &
\colhead{(km/s)} &
\colhead{(km/s)} &
\colhead{(m\AA)} &
\colhead{($10^{12}$ cm$^{-2}$)} &
\colhead{(km/s)} &
\colhead{(km/s)} &
\colhead{(m\AA)} &
\colhead{($10^{12}$ cm$^{-2}$)}
}

\startdata

$R$(0) & 4.96 $\pm$ 0.60 & 4.38 $\pm$ 0.53 & 6.7 & 2.8 & 
         6.87 $\pm$ 0.46 & 6.06 $\pm$ 0.41 & 12.6 & 1.7 & 
        11.83 $\pm$ 0.76 & 10.4 $\pm$ 0.7 \\
$Q$(2) & 9.91 $\pm$ 0.83 & 17.5 $\pm$ 1.5 & 6.7 & 3.1 & 
        10.67 $\pm$ 0.70 & 18.8 $\pm$ 1.2 & 12.4 & 2.2 & 
        20.58 $\pm$ 1.09 & 36.3 $\pm$ 1.9 \\
$Q$(4) & 8.97 $\pm$ 0.75 & 15.8 $\pm$ 1.3 & 6.6 & 3.2 & 
         9.65 $\pm$ 0.59 & 17.0 $\pm$ 1.0 & 12.3 & 2.0 & 
        18.62 $\pm$ 0.95 & 32.8 $\pm$ 1.7 \\
$P$(2) & 2.05 $\pm$ 0.60 & --- & 6.8 & 2.0 & 
         3.38 $\pm$ 0.55 & --- & 12.6 & 1.7 & 
         5.43 $\pm$ 0.81 & --- \\
$Q$(6) & 7.40 $\pm$ 0.66 & 13.0 $\pm$ 1.2 & 6.6 & 3.1 & 
         7.33 $\pm$ 0.53 & 12.9 $\pm$ 0.9 & 12.2 & 2.0 & 
        14.73 $\pm$ 0.85 & 25.9 $\pm$ 1.5 \\
$\left\{\!\! \begin{array}{c} Q(8) \\ P(4) \end{array}\!\! \right\}$ &
--- & --- & --- & --- & --- & --- & --- & --- & 19 $\pm$ 2 \tablenotemark{b} & ---  \\
$P$(4)\tablenotemark{c} & 
--- & --- & --- & --- & --- & --- & --- & --- & 6 $\pm$ 0.3 & ---  \\
$Q$(8)\tablenotemark{d} & 
--- & --- & --- & --- & --- & --- & --- & --- & 13 $\pm$ 2 \tablenotemark{b} & 22.3 $\pm$ 5.2  \\
$Q$(10) & 3.9 $\pm$ 0.6 & 6.6 $\pm$ 1.2 & (6.7)\tablenotemark{e} 
& (3)\tablenotemark{e} & 
          3.0 $\pm$ 0.5 & 5.3 $\pm$ 1.5 & (12.5)\tablenotemark{e}
 & (2.1)\tablenotemark{e} & 
          6.9 $\pm$ 0.8 & 11.9 $\pm$ 2.0  \\
\enddata

\tablenotetext{a}{Statistical uncertainties (1$\sigma$) are listed unless otherwise noted.
Systematic errors are difficult to estimate and may be larger.}

\tablenotetext{b}{Estimate of systematic uncertainty.}

\tablenotetext{c}{Estimated from $Q$(4).}

\tablenotetext{d}{Blend equivalent width minus $P$(4) estimate.}

\tablenotetext{e}{Values constrained in the fit to the ratioed spectrum.}

\end{deluxetable}

\clearpage

\begin{table}
\begin{center}
\caption{C$_2$ Line Parameters for Cygnus OB2 5.\label{C25table}}
{
\footnotesize
\begin{tabular}{ccccccccccc}
Line  & W$_{\lambda}$\tablenotemark{a} & N($J$) \\
& (m\AA) & ($10^{12}$ cm$^{-2}$) \\ \tableline

$R$(0) & 6.6 $\pm$ 0.8 & 5.8 $\pm$ 0.7 \\
$Q$(2) & 10.7 $\pm$ 1.0 & 18.9 $\pm$ 1.8 \\
$Q$(4) & 13.5 $\pm$ 1.0 & 23.8 $\pm$ 1.8 \\
$Q$(6) & 8.0 $\pm$ 1.2 & 14.1 $\pm$ 2.1 \\
$\left\{\!\! \begin{array}{c} Q(8) \\ P(4) \end{array}\!\! \right\}$ &
11 $\pm$ 3\tablenotemark{b} & ---  \\
$P$(4)\tablenotemark{c} & 
4.5 $\pm$ 0.3 & ---  \\
$Q$(8)\tablenotemark{d} & 
7 $\pm$ 3\tablenotemark{b} & 12.3 $\pm$ 5.3  \\
$Q$(10)\tablenotemark{e} & 1.5 $\pm$ 0.5 & 2.6 $\pm$ 0.9  \\
\end{tabular}

\mbox{}

\tablenotetext{a}{Statistical uncertainties (1$\sigma$) are listed unless otherwise noted.
Systematic errors are difficult to estimate and may be larger.}

\tablenotetext{b}{Estimate of systematic uncertainty.}

\tablenotetext{c}{Estimated from $Q$(4).}

\tablenotetext{d}{Blend equivalent width minus $P$(4) estimate.}

\tablenotetext{e}{Marginal feature; values very uncertain.}

}
\end{center}
\end{table}

\clearpage

\begin{table}
\begin{center}
\caption{Line Parameters for CH, CH$^+$\tablenotemark{a}, and CN\tablenotemark{b}.\label{vistable}}
{
\footnotesize
\begin{tabular}{ccccccc}
Object & Species & Line & v$_{\rm LSR}$ & FWHM & W$_{\lambda}$ \tablenotemark{c} & N \tablenotemark{c} \\
       &         &      &  (km/s)       & (km/s) & (m\AA) & ($10^{12}$ cm$^{-2}$) \\ \tableline

Cyg OB2 5 & CH & 4300 & 5.6 & 6.0 & 26.5 $\pm$ 0.7 & 44.0 $\pm$ 1.2 \tablenotemark{d} \\
          &    &      & 12.7 & 3.4 & 7.1 $\pm$ 0.5 & 9.2 $\pm$ 0.6 \tablenotemark{d} \\
HD 183143 & CH & 4300 & 7.7 & 7.7 & 12.3 $\pm$ 0.7 & 17.0 $\pm$ 1.0 \tablenotemark{d} \\
          &    &      & 23.6 & 4.9 & 18.6 $\pm$ 0.6 & 27.7 $\pm$ 0.9 \tablenotemark{d} \\
Cyg OB2 5 & CH$^+$ & 4233 & 2.6 & 5.0 & 11.2 $\pm$ 0.6 & 12.8 $\pm$ 0.7 \\
          &        &      & 7.6 & 4.1 & 7.1 $\pm$ 0.6 & 8.1 $\pm$ 0.7 \\
          &        &      & 13.3 & 8.9 & 11.8 $\pm$ 0.8 & 13.5 $\pm$ 0.9 \\
HD 183143 & CH$^+$ & 4233 & 9.0 & 4.9 & 18.3 $\pm$ 0.5 & 21.0 $\pm$ 0.6 \\
          &        &      & 24.8 & 4.5 & 22.1 $\pm$ 0.5 & 25.3 $\pm$ 0.6 \\
Cyg OB2 12 & CN & A-X R$_2$(0) & 6.8 & 1.1 & 1.3 $\pm$ 0.4 & 3.1 $\pm$ 1.0 \\
           &    &              & 12.4 & 1.7 & 4.3 $\pm$ 0.6 & 10.3 $\pm$ 1.4 \\
Cyg OB2 5  & CN & A-X R$_2$(0) & 5.6 & 2.0 & 1.2 $\pm$ 0.3 & 2.9 $\pm$ 0.7 \\
           &    &              & 13.7 & 1.7 & 1.4 $\pm$ 0.3 & 3.4 $\pm$ 0.7 \\
           &    & B-X R(0) & 5.6 & 3.9 & 4.1\tablenotemark{e} & 0.9\tablenotemark{e} \\
           &    &          & 12.1 & 3.9 & 2.8\tablenotemark{e} & 0.6\tablenotemark{e} \\
HD 183143  & CN & B-X R(0) & 25.2 & 2.5 & 5.6 $\pm$ 0.4 & 1.3 $\pm$ 0.1 \\
           &    & B-X R(1) & 24.1 & 1.5 & 1.2 $\pm$ 0.2 & 0.4 $\pm$ 0.1 \\

\end{tabular}

\tablenotetext{a}{For CH$^+$, we adopt the oscillator strength 
$f_{00} = 5.45 \times 10^{-3}$ of \citet{larsson1}.}

\tablenotetext{b}{For CN, we adopt the oscillator strengths
$f^{\rm BX}_{00} = 3.38 \times 10^{-2}$ \citep{larsson2}
and $f^{\rm AX}_{20} = 7.6 \times 10^{-4}$ \citep{davis}.}

\tablenotetext{c}{Statistical uncertainties (1$\sigma$) are listed unless otherwise noted.
Systematic errors are difficult to estimate and may be larger.}

\tablenotetext{d}{Estimated using the CH curve of growth of 
\citet{EvD:CH}, assuming $b$=1 km/s.  All other column densities computed
assuming the lines are optically thin.}

\tablenotetext{e}{These values are extremely uncertain due to the low flux of Cygnus OB2 5
at violet wavelengths.}

}

\end{center}
\end{table}

\clearpage

\begin{table}
\begin{center}
\caption{Cloud Parameters Based on Conventional Chemical Model.\label{parameters}}
{
\footnotesize
\begin{tabular}{ccccccc}
Object & $N$(\hhh) & L & d & E(B-V) & $N$(H$_2$) + $N$(H) & $\langle n \rangle$ \\
 & ($10^{14}$ cm$^{-2}$) & (pc) & (pc) & (mag) & ($10^{21}$ cm$^{-2}$) & (cm$^{-3}$) \\ \tableline

Cyg OB2 12 & 3.8\tablenotemark{a} & 905 & 1700\tablenotemark{b} & 3.35\tablenotemark{c} & 12.95 & 4.8 \\ 
Cyg OB2 5  & 2.6\tablenotemark{d} & 629 & 1700\tablenotemark{b} & 1.99\tablenotemark{c} & 7.69 & 4.1 \\
HD 183143  & 2.3\tablenotemark{d} & 552 & 1000\tablenotemark{e} & 1.28\tablenotemark{f} & 4.95 & 3.0 \\
HD 20041   & 3.5\tablenotemark{g} & 833 & 1400\tablenotemark{e} & 0.70\tablenotemark{h} & 2.71 & 1.1 \\
WR 121     & 2.2\tablenotemark{g} & 524 & 1690\tablenotemark{i} & 1.68\tablenotemark{j} & 6.50 & 4.1 \\
WR 104     & 2.3\tablenotemark{d} & 547 & 1300\tablenotemark{i} & 2.10\tablenotemark{j} & 8.12 & 4.9 \\
WR 118     & 6.5\tablenotemark{d} & 1548 & 6300\tablenotemark{i} & 4.13\tablenotemark{j} & 15.97 & 3.4 \\
HD 194279  & $<$1.2\tablenotemark{g} & $<$286 & 1100\tablenotemark{e} & 1.22\tablenotemark{f} & 4.72 & $>$5.5 \\
HD 168607  & $<$0.6 & $<$143 & 1100\tablenotemark{e} & 1.61\tablenotemark{f} & 6.23 & $>$14.5 \\
P Cygni    & $<$0.6\tablenotemark{g} & $<$143 & 1800\tablenotemark{k} & 0.63\tablenotemark{k} & 2.44 & $>$5.7 \\
$\chi^2$ Ori & $<$0.7\tablenotemark{g} & $<$171 & 1000\tablenotemark{e} & 0.44\tablenotemark{l} & 1.70 & $>$3.3  \\
$\zeta$ Oph & $<$0.4\tablenotemark{g} & $<$86 & 140\tablenotemark{m} & 0.32\tablenotemark{f} & 1.24 & $>$4.8 \\

\end{tabular}

\mbox{}

\tablenotetext{a}{From \citet{mccalldiffuse}.}

\tablenotetext{b}{\citet{cygdist}.}

\tablenotetext{c}{From \citet{schulte}.}

\tablenotetext{d}{The UKIRT observations of $R$(1,0) and $R$(1,1)$^u$ are adopted for
these sources.}

\tablenotetext{e}{Our estimate (spectroscopic parallax).}

\tablenotetext{f}{From \citet{snowyorkwelty}.}

\tablenotetext{g}{Total \hhh\ column density estimated to be twice that of {\em para}-\hhh.}

\tablenotetext{h}{\citet{racine}.}

\tablenotetext{i}{\citet{wolfrayet}.}

\tablenotetext{j}{From \citet{pendleton}, assuming 
$R_V$ = $A_V$ / E(B-V) = 3.1.}

\tablenotetext{k}{From \citet{pcygni}.}

\tablenotetext{l}{From the intrinsic color of \citet{wegner}.}

\tablenotetext{m}{From Hipparcos catalog \citep{hipparcos}.}

}
\end{center}
\end{table}

\clearpage

\end{document}